\documentclass[a4paper,11pt]{article}
\usepackage{aaskaiid}
\usepackage{xcolor}
\usepackage{comment}
\usepackage{orcidlink}

\newcommand{\Rsun}{R$_\odot$}

\title{The Study of Quasi-Periodic Pulsations in Solar and Stellar Flares with SKA}
\ShortTitle{QPPs in solar and stellar flares}

\author[1]{Valery M. Nakariakov\orcidlink{0000-0001-6423-8286}}
\emailAdd{V.Nakariakov@warwick.ac.uk}
\author[2,3]{Atul Mohan\orcidlink{0000-0002-1571-7931}}
\emailAdd{atul.multiverse73@gmail.com}
\author[4,5]{Vishal Upendran\orcidlink{0000-0002-9253-6093}}
\author[3,10]{Teresa Monsue\orcidlink{0000-0003-3896-3059}}
\author[6]{Hamish A.S. Reid\orcidlink{0000-0002-6287-3494}}
\author[1,7] {Dmitrii Y. Kolotkov\orcidlink{0000-0002-0687-6172}}
\author[1] {Sergey A. Belov\orcidlink{0000-0002-3505-9542}}
\author[8,9]{Kyung-Suk Cho\orcidlink{0000-0003-2161-9606}}
\affiliation[1]{Centre for Fusion, Space and Astrophysics, Physics Department, University of Warwick, Coventry CV4 7AL, UK}
\affiliation[2]{Solar Physics Laboratory, NASA Goddard Space Flight Center Greenbelt, MD 20771, USA}
\affiliation[3]{Department of Physics, The Catholic University of America, 620 Michigan Ave NE, Washington, DC 20064}
\affiliation[4]{SETI Institute, Mountain View, CA, USA - 94043}
\affiliation[5]{Lockheed Martin Solar and Astrophysics Laboratory, Palo Alto, CA, USA - 94306}
\affiliation[6]{Mullard Space Science Laboratory, University College London, Holmbury St. Mary, Dorking, Surrey, RH5 6NT, UK}
\affiliation[7]{Engineering Research Institute \lq\lq Ventspils International Radio Astronomy Centre (VIRAC)\rq\rq, Ventspils University of Applied Sciences, Ventspils, LV-3601, Latvia}
\affiliation[8]{Space Science Division, Korea Astronomy and Space Science Institute, Daejeon 305-348, Republic of Korea} 
\affiliation[9]{Department of Astronomy and Space Science, University of Science and Technology, Daejeon 305-348, Republic of Korea}
\affiliation[10]{CRESST II and Exoplanets and Stellar Astrophysics Laboratory, NASA Goddard Space Flight Center Greenbelt, MD 20771, USA}
\abstract{
    An intensively studied phenomenon which is not described by the standard flare model are quasi-periodic pulsations (QPP) of the flaring emission. As analysis of the QPP phenomenon intrinsically requires a combination of high time and spatial resolutions, especially in the radio band, the unprecedented capabilities of SKA offer us a unique opportunity to reach a breakthrough progress in the observational study of QPP. The SKA-Mid-frequency band falls in a unique window where both coherent emissions from particle acceleration sites and incoherent gyrosynchrotron emissions from non-thermal particles in coronal loops can be studied. With an additional polarisation dimension and the capability to perform wideband spectroscopic imaging, the QPPs in gyrosynchrotron emission ($\ge 1$~GHz) and plasma emission will help understand the local magnetic field modulation due to active phenomena and the response seen in the particle acceleration observable below ~600~MHz. An incomplete list of specific science questions to be addressed with SKA includes (a) the role of QPP in the energy partition in flares, (b) seismology of flaring sites by QPP of different classes, (c ) differences and similarities between QPP in solar and stellar flares, (d) advancing the standard flare model, (e) the physics of repetitive magnetic reconnection: spontaneous vs induced, (f) ML techniques in the detection, classification and analysis of QPP, (g) QPP in weak flares. The latter topic could be especially advanced with SKA which will allow for high-cadence high fidelity radio imaging of weak energy release events.  }
\begin{document}
\maketitle

\section{Introduction}
\label{sec:intro}
\begin{figure}[htb]
  \centering
  \includegraphics[width=0.35\textwidth]{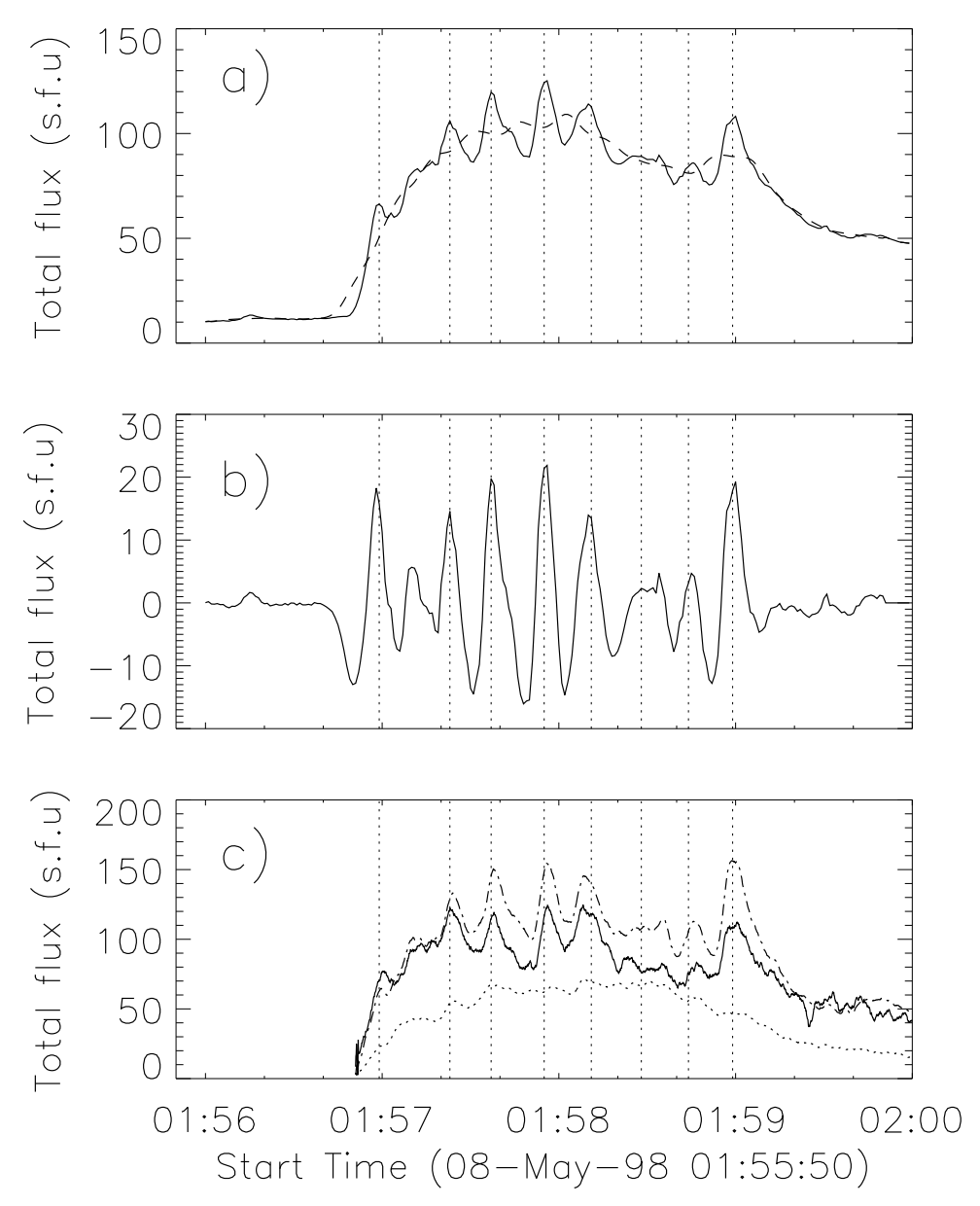}
  \includegraphics[width=0.5\textwidth]{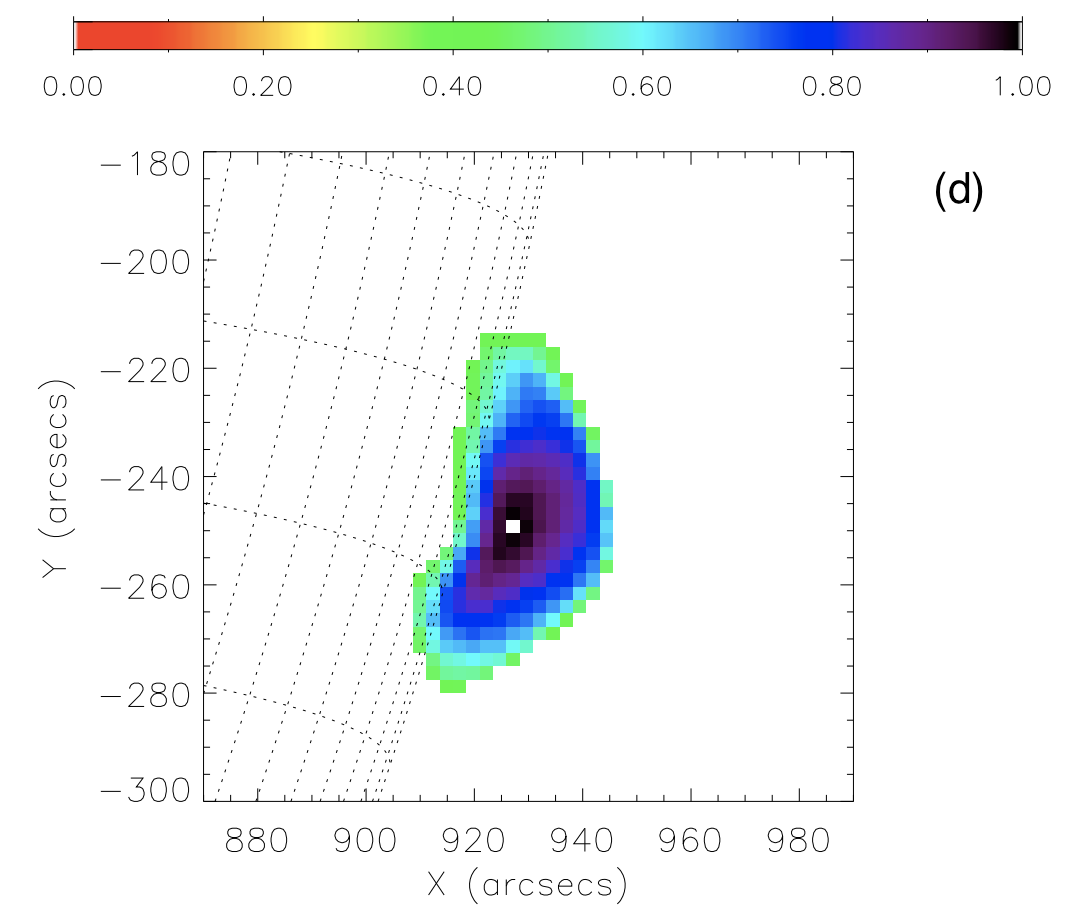}
  \caption{(a) Integrated flux time profile of microwave emission at 17~GHz for an off-limb solar flare on 8th May 1998, as observed by the Nobeyama Radioheliograph. Overlayed
with the dashed line is the background emission profile obtained via 20~s
smoothing of the integrated flux signal. b) The signal after subtraction
of the background profile from the integrated flux. c) Solar flux time
profiles of microwave emission at 9.4~GHz (dot-dashed line), 17~GHz
(solid line) and 3.75~GHz (dotted line), obtained by the Nobeyama
Radiopolarimeters.
(d) The cross-correlation coefficient map for background subtracted microwave emission at 17~GHz, obtained by the Nobeyama Radioheliograph during the QPP. The correlation coefficient is calculated with zero time lag. The master pixel is the pixel with the highest time-integrated intensity of the signal. 
Figure credits: \cite{2008A&A...487.1147I}.}
   \label{fig00}
\end{figure}

The phenomenon of quasi-periodic pulsations (QPP) in solar flares has been attracting attention for several decades, since the discovery of 16-s periodic modulation of X-ray intensity of a flaring emission, detected on a high-altitude balloon \citep{1969ApJ...155L.117P}. QPP appear in all types of the emitted radiation, both thermal and non-thermal, from radio and millimeter to X-ray and gamma-ray wavelengths \citep[see, e.g.,][for reviews]{2009SSRv..149..119N, 2010PPCF...52l4009N, Krucker13_mmSolarflareRev, 2016SoPh..291.3143V, 2020STP.....6a...3K, 2021SSRv..217...66Z}. Typical oscillation periods range from a fraction of a second to several tens of minutes. An example of a spatially resolved QPP in the microwave emission of an an M-class solar flare is shown in Figure~\ref{fig00}. 
QPP are detected in all classes of solar flares, from microflares to the most powerful X-class flares, and in all phases of a flare. Statistical analyses based upon rather conservative detection criteria demonstrate that QPP is a common feature of solar flares \citep{2020ApJ...895...50H, 2015SoPh..290.3625S}. Specific attention has recently been given to non-stationary QPP patterns, with pronounced modulation of the instantaneous oscillation period and amplitude, as the modulations could reveal evolution of parameters of flaring cites \citep{2019PPCF...61a4024N}.  

Similar QPP patterns have been detected in lightcurves of stellar flares, in radio \cite[e.g.,][]{2001A&A...374.1072S}, UV \citep[e.g.,][]{2018MNRAS.475.2842D}, and white light \citep[e.g.,][]{2016MNRAS.459.3659P, 2020A&A...636A..96M, 2022MNRAS.514.5178D} emissions, including powerful superflares.
Moreover, a similarity of QPPs, at least of a certain class, in stellar and solar flares has been established empirically: a statistically similar linear scaling of the damping time with the oscillation period of QPPs of the SUMER-oscillation class (a rapidly decaying almost harmonic oscillations of the thermal emission intensity in the decay phase of the flare) \citep{2016ApJ...830..110C}. This analogy opens up promising opportunities for a comparative analysis of QPPs in stellar and solar flares and their relationship with flare parameters. It substantially broadens the accessible parameter space, which is essential for understanding the connection between solar flares and stellar superflares, the processes responsible for generating QPPs, and the seismological diagnostics of plasma in flare regions.

The broad variety of oscillation periods (ranging from sub-seconds to almost hours), modulations, and kinds of modulated emissions suggests that different types of QPP have different nature. Theoretical mechanisms which have been proposed to explain the phenomenon of QPP could be divided into three main groups: the modulation of the parameters of the emitting plasma by an external periodic process, for example, magnetohydrodynamic (MHD) oscillations; and repetitive magnetic reconnection which is either spontaneous, i.e., a self-oscillation, or is induced by an external driver, see \citep{Asch_QPPtheoryRev1987,2018SSRv..214...45M, 2021SSRv..217...66Z} for comprehensive reviews. Mechanisms based on repetitive reconnection include the periodic acceleration of non-thermal charged particles. In addition, the kinematics of particles accelerated impulsively, randomly, or steadily can be periodically modulated by MHD oscillations, resulting in QPPs. Although QPPs are a widespread feature of flaring energy release, the current form of the standard flare model does not account for this phenomenon. Therefore, incorporating QPPs into the model offers a promising path toward a more reliable understanding of the flare process, including its forecasting.

One of the bottlenecks in distinguishing the QPP mechanisms is the ability to image the particle acceleration sites and study their variability at sub-arcsecond resolution across the active region as the event progresses.
Radio observations are extremely sensitive to reconnection-driven electron beams, owing to coherent emission mechanisms involving wave-wave and wave-particle resonant interactions triggered by electron beams~\citep{ginzburg1958,Tsytovich69,melrose1970,melrose1972}. Hence, even weak reconnection events can generate radio bursts with emission brightness temperatures up to $10^9$--10$^{11}$\,K~\citep{Hilaire13_typeIIIstats,Reid2014}. Also, since radio burst emission forms as coherent emission at the local plasma or gyro- frequency ($\nu_\mathrm{p}$ or $\nu_\mathrm{B}$) and its harmonics due the plasma emission and electron cyclotron maser mechanisms, the observation frequency directly links to local density or magnetic field strength at emission sites. Hence, broadband spectroscopic imaging at high time resolution can map the variability at different regions simultaneously and directly infer the evolution of local physical parameters at the QPP source.

Furthermore, imaging-based studies could hardly explore the evolution of the pre- and post-flare sources  because of limitations in (a) imaging sensitivity and (b) dynamic range required for the robust detection and morphological characterization of the source. The latter issue is often a serious problem since there can be multiple co-temporal active regions with coherent radio flux levels varying by 2--4 orders of magnitude at fine time-frequency scales on the solar disk. 
This problem limited the extension of imaging-based studies to the larger ensemble of weak solar flares, which play a major role in quasi-steady heating and particle acceleration in solar active regions~\citep{ash2012_flarestats}.
Weak flares, which are believed to be phenomenologically similar to strong events, also present QPP patterns~\citep[e.g.,][]{2018ApJ...859..154N}. 
The SKA-Low and mid arrays in AA* and AA4 configuration will enable sub-arcsec scale high dynamic range imaging capability with sub-second, sub-MHz spectro-temporal averaging, revolutionizing the exploration of radio QPP sources associated with a range of flare energy scales from nanoflares to major eruptions.
Combining this unique radio data with high-cadence optical-to-X-ray spectroscopy data from cutting-edge ground- and space-based instruments will enable us to tackle the major open problems in the field~\citep{2015aska.confE.169N}.
An incomplete list of specific science questions to be addressed includes (a) the role of QPPs in the energy partition in flares, (b) seismology of flaring sites by QPPs of different classes, (c) advancing the standard flare model, (d) the physics of repetitive magnetic reconnection: spontaneous vs induced, (e) ML techniques in the detection, classification and analysis of QPP, and (f) QPP in nanoflares.
Besides, a statistical understanding of the link between various classes of QPPs, underlying activity, and associated physical models will help advance our understanding of QPPs observed in sun-like stars, observations of which lack spatially resolved information.

The state-of-the-art SKA precursors and pathfinders primarily in the low frequency band (0.05--0.35\,GHz) have already enabled high dynamic range snapshot spectroscopic imaging of solar active regions and quasi-steady emission regions across corona at sub-second and sub-MHz resolution.
These instruments include the SKA precursor, Murchison Widefield Array \citep[MWA;][]{Tingay2013,Wayth18_MWA_phaseII}, operating in the 80--240\,MHz and the LOw Frequency ARray~\citep[LOFAR;][]{vanHaarlem13_LOFAR} in the 30--80\,MHz band, covering the planned SKA-Low spectral band.

In this chapter, we briefly summarise mechanisms that are considered to be responsible for QPP (Sec.~\ref{sec:mech}); address the potential of the QPP study to reveal the processes behind stellar flares (Sec.~\ref{sec:stell}); describe promising data analysis 
techniques relevant to the detection and analysis of QPP in observational data (Sec.~\ref{sec:techs}) and the machine learing approach (Sec.~\ref{sec:ml}; overview advances in QPP research with SKA precursors and pathfinders (Sec.~\ref{sec:prec}); discuss expected advances with SKA-Low and SKA-Mid (Sec.~\ref{sec:exp}), including possibilities for coordinated observation with other upcoming observational facilities. Bried conclusions are given in Sec.~\ref{sec:conc}.

\section{Mechanisms for QPP}
\label{sec:mech}

The use of QPP for the diagnostics of flaring plasmas, to understand the underlying physical processes, and to exploit the solar–stellar analogy requires identification of the mechanisms responsible for the QPP phenomenon. The wide range of observed periods and the diversity of modulation patterns strongly suggest that multiple physical processes can produce signals that appear as QPP. Currently, there is no consensus on which specific mechanisms are responsible for a given type of QPP.
The recent comprehensive review by \citet{2021SSRv..217...66Z} summarises fifteen distinct mechanisms capable of generating QPP. In this section, we briefly summarise those mechanisms that are most relevant for producing QPP detectable with SKA.

\subsection{MHD oscillations in coronal plasma structures}

A major class of explanations attributes QPP to MHD oscillations of plasma structures. Such oscillations, primarily fast and slow magnetoacoustic modes, are detected confidently in the corona \citep[e.g.,][]{2020ARA&A..58..441N}. They can be excited by flare-associated impulsive drivers in standing or propagating forms. Once excited, these waves modulate local plasma parameters, especially density, and the strength and direction of the magnetic field. The resulting variation of the emitting plasma produces observable modulation in both thermal and non-thermal emissions.

Fast sausage modes which are axisymmetric perturbations of, for example, coronal loops, are highly compressive and can modulate density and magnetic field strength efficiently, making them strong candidates for QPP in the microwave emission \citep[e.g.,][]{2005A&A...439..727M, 2015SoPh..290.1173K}, and also the coherent radio emission by perturbing the local electron plasma frequency. Oscillation periods of sausage modes range from sub-second to tens of seconds depending on loop radius, density contrast, and the magnetic field strength. 

Fast kink modes, although only weakly compressive in the long-wavelength limit, cause periodic variations in the direction of the local magnetic field and can therefore modulate the intensity of microwave emission through changes in the column depth of the emitting region. In addition, QPP modulation arises from periodic variations of the angle between the magnetic field and the line of sight. The periods of kink oscillations are determined by the internal and external Alfvén speeds of the oscillating plasma structure (e.g. a coronal loop) and by its length. Typical kink-mode periods range from tens of seconds to several tens of minutes.

Alfvén waves may also modulate gyrosynchrotron emission by altering the local angle between the magnetic field and the line of sight. However, the essentially non-collective nature of coronal Alfvén waves prevents them from producing systematic QPP.

Slow magnetoacoustic modes produce QPP by modulating the plasma density and hence the local electron plasma frequency. 
Typical periods of standing slow oscillations detected in the corona range between a few and tens of minutes, while periods of propagating slow waves, about 3 minutes, are prescribed by the chromospheric driver.

\subsection{Periodic triggering of magnetic reconnection by external waves}

Rather than oscillations of the emitting plasma itself, QPP may also arise from the periodic modulation of magnetic reconnection by external waves, e.g., by slow \citep{2006SoPh..238..313C} and  fast \citep{2006A&A...452..343N}. 
The waves cause variations in local plasma parameters such as the magnetic field and density in the vicinity of the reconnection site.  Furthermore, from the point of view of the reconnection site, an incoming perpendicular longitudinal wave, i.e., a compressive wave propagating across the field, causes the variations of the inflow rate. 
These variations affect the reconnection rate and hence the effectiveness of the instantaneous energy release, including acceleration of non-thermal electorns. The oscillation periods of periodically triggered QPP are determined by the time variability in the triggering wave motion.  

In addition, magnetoacoustic waves approaching the reconnection site experience refraction toward magnetic null points
\citep{2011SSRv..158..205M}, where they concentrate energy and produce spikes of the elecric current density. When such waves arrive periodically, they can generate periodic episodes of the current-diven anomalous electrical resistivity and
hence bursts of reconnection \citep{2006A&A...452..343N}. Even small-amplitude waves can lead to strong current modulation because the null region acts as a geometric amplifier (a lense, focusing the waves in its vicinity) and because of the highly nonlinear nature of the electrical resistivity dependence of the current density.

\subsection{Spontaneous oscillatory reconnection}

A fundamentally different class of mechanisms concerns repeatitive or {oscillatory reconnection}, in which reconnection proceeds in a self-sustained, quasi-periodic manner even when the driver is not periodic
\citep[e.g.,][]{2012A&A...548A..98M, 2019A&A...621A.106T, 2019PhPl...26k2110M}. This behaviour emerges during the relaxation of stressed magnetic fields containing null points or current sheets, and the dynamic interplay between Lorentz forces and thermal-pressure gradients. 

For example, in one of the proposed 2D scenarios, an initial collapse of the reconnection site forms a current sheet and launches outflows that heat the surrounding plasma. The heated plasma then expands, overcompresses the null region, and forces the system into a current sheet of opposite orientation. This sequence repeats several times, producing bursts of reconnection with alternating inflow and outflow geometry \citep{2009A&A...493..227M}. Each cycle dissipates a fraction of the stored magnetic energy, leading to exponentially decaying amplitudes. QPP produced by this mechanism may appear in all observational bands.

There has not been a systematic study that would link the oscillation period with parameters of the reconnection site and its vicinity, while various case studies show a broad range of oscillation periods. In particular, numerical simulations of collapse of a single magnetic null point yield characteristic periods of 50--80~s \citep{2012A&A...548A..98M}, while flux-emergence scenarios generate periods of 100--200~s \citep{2009A&A...494..329M}.
According to a numerical study of \citet{2023ApJ...943..131K}, in a 1~MK plasma, the oscillation period of repeative reconnection in a 2D magnetic null point without a guiding field is estimated as 
$P_{\mathrm{rr}} \approx 38.0/B_{\mathrm{0}} + 26.08 \times 10^{6} \sqrt{\rho_{\mathrm{0}}}  - 21.01 \pm 4.26$,
where the period $P_{\mathrm{rr}}$ is measured in seconds, the plasma density $\rho_{\mathrm{0}}$ in kilograms per cubic meters, and $B_{\mathrm{0}}$ is the magnetic field at a distance 1 Mm from the null point in Gauss.

\subsection{Self-oscillatory and over-stable plasma processes}

Similarly to spontaneous repetitive reconnection, acoustic or magnetoacoustic self-oscillatory processes may arise when (quasi-)steady energy input (heating or flows) competes with losses (radiation, conduction, viscosity, and resistivity) resulting in an oscillatory variation of the plasma parameters. 
One of the examples is the phenomenon of thermal over-stability. In the corona, thermal over-stability is characterised by a repeatitive variation of thermal equilibrium established by the balance of optically thin radiative cooling, cooling by field-aligned thermal conduction down to the chromosphere, and unspecified plasma heating mechanism.  The oscillation period of thermal over-stability is determined the local sound speed, i.e., the plasma temperature, and derivatives of the combined heating/cooling function with respect to local thermodynamic parameters, e.g., denisty and temperature. 
Perturbations of the plasma density and temperature, caused by thermal instability, lead to the variation of thermal emission from the plasma, and can also affect radio and microwave emission by the variation of the electron plasma frequency.

Flow-driven over-stability is associated with steady inflows into a reconnecting current sheet. When flow speeds approach the Alfv\'en speed, coupled Kelvin--Helmholtz and tearing instabilities generate over-stable oscillations \citep{2006ApJ...644L.149O}. These cause periodic variability in the Ohmic heating rate, plasmoid formation, and bursts of particle acceleration, and hence can produce QPP in both thermal and non-thermal emissions.

In fragmented current sheets, recurrent plasmoid ejection can itself generate quasi-periodic energy release by the \lq\lq magnetic tuning fork\rq\rq\ mechanism  \citep{2016ApJ...823..150T}. In this mechanism, a pair of repeatedly diverging and converging magnetic surfaces near a reconnection region behave like the prongs of a tuning fork, producing quasi-periodic compressions and rarefactions of the plasma. In this scenario, fast reconnection outflows collide with a dense magnetic arcade or loop system; the reflected back-pressure periodically squeezes and re-expands the outflow channel, creating an oscillatory pattern in the current sheet and surrounding magnetic field. This leads to periodic acceleration or injection of energetic electrons, and hence to quasi-periodic modulation of microwave and hard X-ray emission. The oscillation period is set by the local Alfvén speed and the size of the reconnection outflow region, typically yielding periods of a few seconds to tens of seconds.

\subsection{Equivalent LCR-contour}

The LCR mechanism for QPP interprets a flaring coronal loop as an equivalent electrical circuit containing inductive (L), capacitive (C), and resistive (R) elements, capable of supporting oscillatory electric currents \citep[e.g.,][]{2008PhyU...51.1123Z}. In the chromospheric segment of the circuit, the current between the footpoints arises because of the finite perpendicular electrical conductivity associated with partial ionisation effects. In this scenario, an impulsive energy release perturbs the alteernate current flowing along the loop and the chromospheric \lq\lq short cut\lq\lq, after which the system relaxes through oscillations governed by the circuit’s natural eigenfrequency.

These current oscillations modulate the gyrosynchrotron emissivity by periodically varying the magnetic field strength, pitch-angle distribution, and density of non-thermal electrons, thereby producing QPP in microwave and hard X-ray emission. In addition, Ohmic dissipation of the alternate current can lead to oscillatory behaviour in the thermal emission. The period of the LCR oscillation depends on the loop length, magnetic flux, plasma density, and effective electrical resistivity in both the coronal and chromospheric parts of the ``circuit'', and typically lies in the sub-second to several-second range, although longer periods are possible in loops with high inductance or low plasma resistivity. 

A limitation of this mechanism is that in a magnetic flux tube, the alternate current is associated with alternate magnetic twisting, i.e. torsional Alfv\'en waves, whose resonant periods predicted by MHD theory are generally an order of magnitude longer. Nevertheless, observations of 1.5-hour QPP in soft X-rays and 3-hour QPP in white-light emission during a stellar flare, detected in thermal and non-thermal channels, respectively, are consistent with an LCR-type oscillatory scenario \citep{2021ApJ...923L..33K}.

Table~\ref{table1} summarises the most widely discussed mechanisms for QPP and provides very approximate characteristic properties of the resulting pulsations.

\begin{table}[ht]
\centering
\caption{Comparison of the principal mechanisms proposed for QPP generation in solar and stellar flares.}
\begin{tabular}{p{3.5cm} p{2.5cm} p{5.8cm} p{3.2cm}}
\hline
\textbf{Mechanism} & \textbf{Typical Periods} & \textbf{Key Indicators} & \textbf{Dominant Channels} \\
\hline

MHD eigenmodes (sausage, kink, slow) of emitting plasma structures
& $~$1~s -- $~$1~hour 
& Harmonic decaying or decayless patterns; periods set by AR geometry; spatial oscillation signatures; multi-period behaviour possible. 
& Thermal (EUV, SXR) and non-thermal (radio, microwave HXR) \\[0.2cm]

Periodic triggering of reconnection by external waves 
& $~$1~s -- $~$1~hour, depending on nature of the external wave 
& Coherence with oscillations in chromospheric or coronal plasma structures. Oscillatory patterns can be unharmoniic 
& Radio, microwave, EUV, SXR, HXR \\[0.2cm]

Spontaneous oscillatory reconnection 
& 50--200~s 
& Sequence of reconnection bursts; anharmonic signals 
& Non-thermal. Thermal for periods $>1$--2~min  \\[0.2cm]

Thermal over-stability 
& Several tens of seconds to many minutes 
& Growing or decayless thermal QPP; strong temperature/density modulation; preflare long-period pulsations. 
& SXR, EUV, radio, microwave \\[0.2cm]

Flow-driven over-stability 
& 100--200~s 
& Variation of Ohmic heating; plasmoid formation; simultaneous thermal and non-thermal modulation. 
& Radio, microwave, EUV, SXR, HXR \\[0.2cm]

Dispersive fast-wave trains
& $<$1~s to $\sim$1--2~min 
& Wave-train shape of amplitude, time variation of period. 
& Radio, microwave, EUV, SXR, HXR \\[0.2cm]

Equivalent LCR-contour 
& $\sim$0.1 -- a few hours 
& Oscillatory alternated current in the flaring loop. Possiibly produces highly regular QPP. For large modulation depths, nonthermal emission oscillation period is two longer than thermal one. 
& Radio, microwave; occasionally HXR \\[0.2cm]

\hline
\label{table1}
\end{tabular}
\label{tab:qpp_mechanisms}
\end{table}

\section{Solar-stellar analogy in the QPP study}
\label{sec:stell}

\begin{figure}[htb]
  \centering
  \includegraphics[width=0.6\textwidth, height=0.4\textwidth]{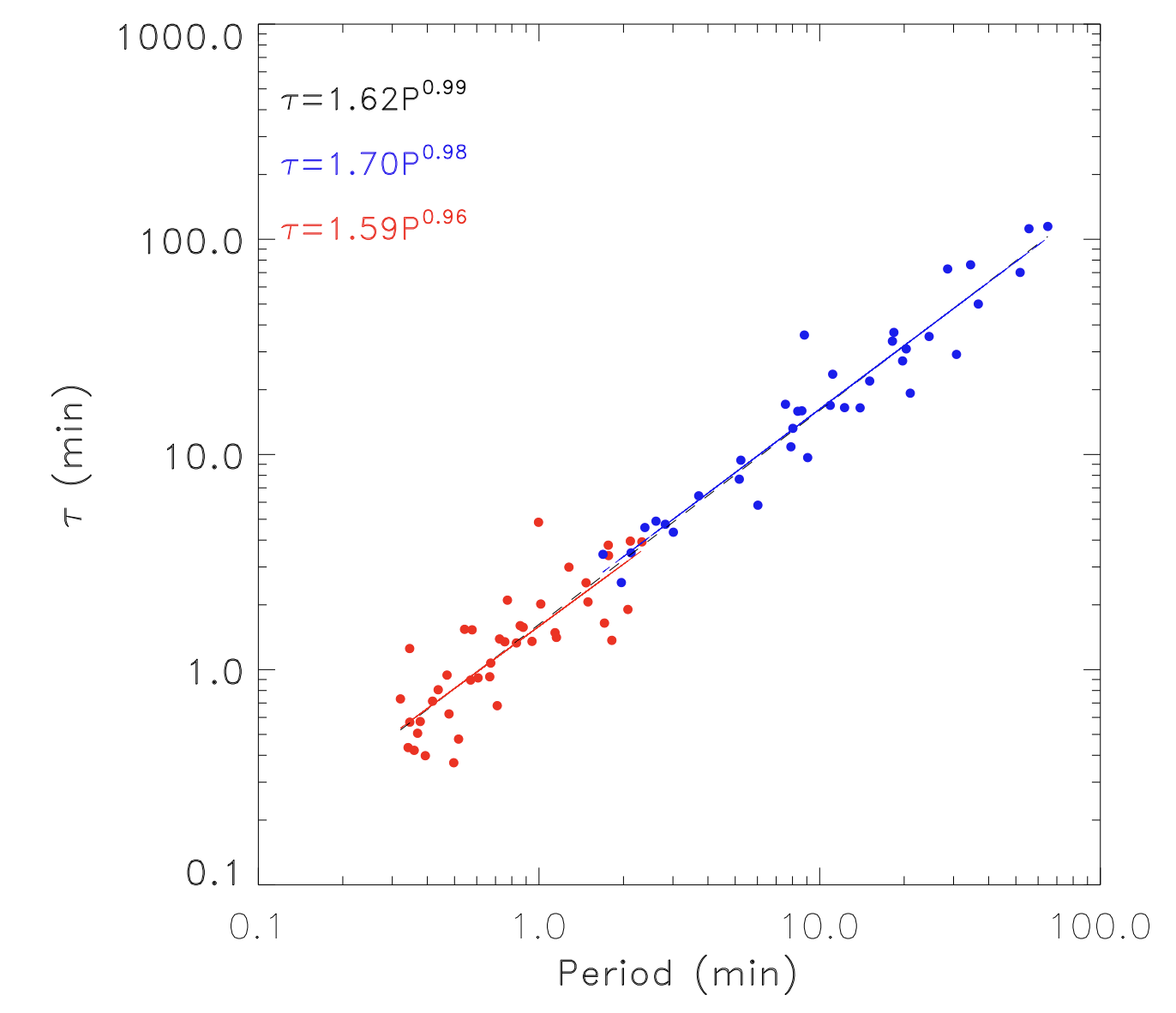}
   \caption{Damping times as a function of the oscillation period for the solar (red) and stellar QPPs (blue) of the SUMER type, detected in the soft X-ray emission in the decay phases of the flares. The blue and red straight lines show the best-fitting powerlaw dependency. The black dashed line the linear fit 
     of the combined, solar and stellar, sets of QPPs. Figure credits:\citep{2016ApJ...830..110C}}
   \label{fig:cho}
   \vspace{-0.5cm}
\end{figure}

Despite growing interest, detections of QPPs in stellar flares remain sporadic. The main hurdles are the limited sensitivity and time resolution of available instruments, as well as the need for long \lq\lq sit-and-stare\rq\rq\ observations to capture a flare during the relevant interval. Recently, most QPP detections in stellar flares have been made in the white-light channel \citep[see, e.g.,][]{2016MNRAS.459.3659P, 2022ApJ...926..204H, 2024A&A...686A.239B, 2025A&A...700A.178J}, using spaceborne telescopes primarily designed for exoplanet searches.

Nevertheless, there is increasing evidence for the similarity of QPPs in solar and stellar flares, at least for certain classes of pulsations. \citet{2016ApJ...830..110C} analysed and compared SUMER-type QPPs detected in soft X-ray emission from both solar and stellar flares. They find that solar QPPs tend to have shorter oscillation periods than their stellar counterparts in the same QPP class, which may be attributed to the generally more powerful nature of the detected stellar flares. However, the damping times and oscillation periods exhibit a linear scaling in both solar and stellar cases (see Figure~\ref{fig:cho}). This suggests that the underlying physical processes responsible for the modulation are similar, providing a foundation for using stellar-flare QPPs as a tool for seismological diagnostics.

\begin{figure}[htb]
 \vspace{-0.1cm}
  \centering
  \includegraphics[width=\textwidth,height=0.38\textheight]{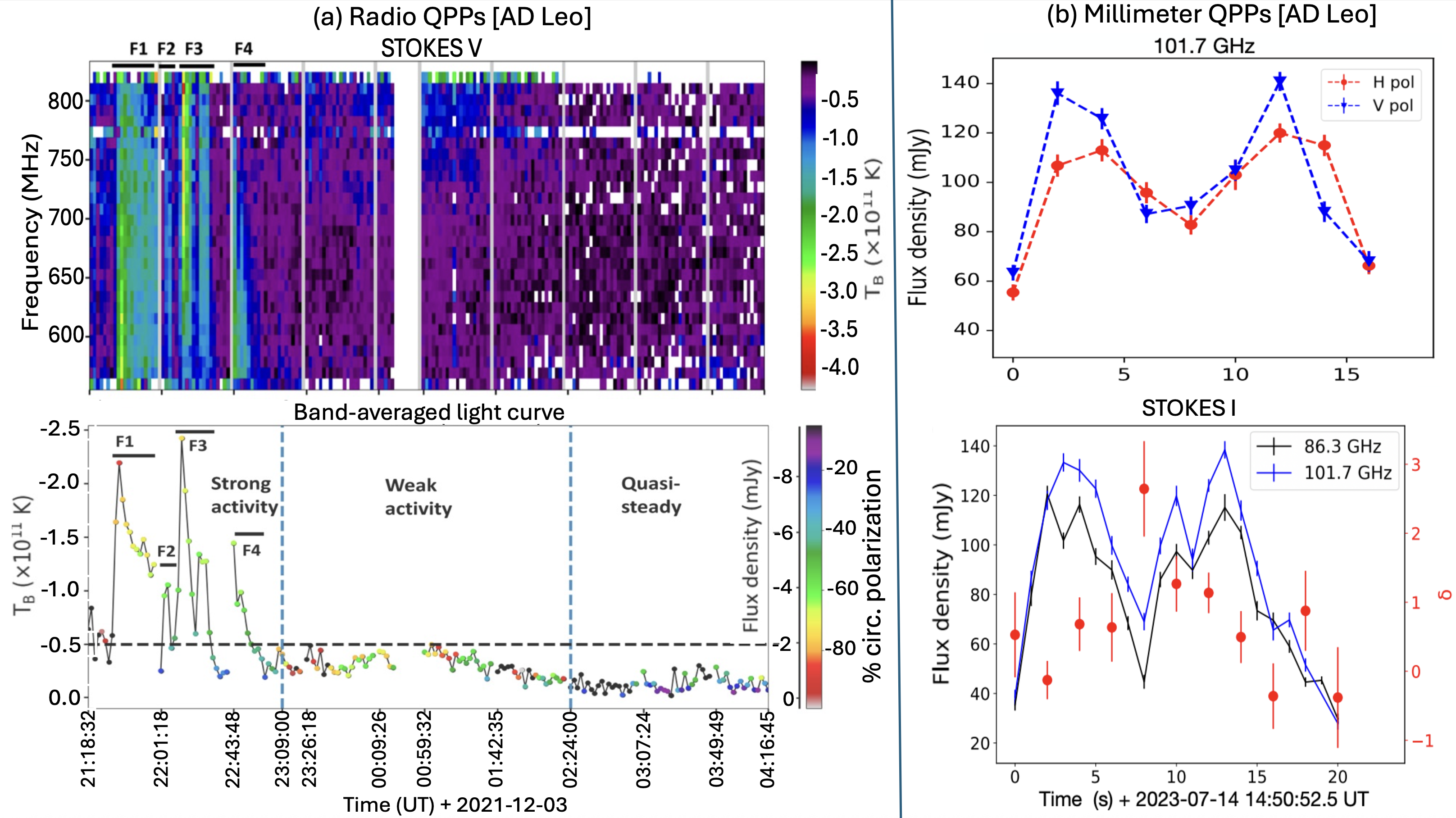}
   \caption{QPPs in AD\,Leo bursts. (a): STOKES V dynamic spectrum and band-averaged light curve showing highly polarised quasi-periodic radio bursts (F1-4) during a strong activity period. The periods of strong and weak activity show type III and band-limited type IV burst features, respectively, marking this the first case to report these burst-types in a young M-dwarf~\citep{Atul24_ADLeotyIV}. Horizontal line in the light curve marks the base level adopted to define strong activity (b): Millimeter bursts recorded by NOEMA showing second-scale QPPs in the two linear polarization channels (H and V). The mean STOKES I light curves in the two sub-bands reveal a frequency-rising nature (positive spectral index ($\delta$)) with second-scale modulations on the $\sim$8-13\,s scale burst pulses~\citep{Atul25_ADLeo_mm}.}
   \label{fig:star_radio-mmQPP}
   \vspace{-0.55cm}
\end{figure}

Detections of QPPs in the radio and microwave emission of stellar flares are much rarer. 
For example, \citet{2004AstL...30..319Z} reported $~$1-s QPPs in the 4.5--5.1~GHz flaring emission of the M-dwarf AD Leo using the 100-m Effelsberg radio telescope. 
More recently, observations of AD Leo with the Five-hundred-meter Aperture Spherical Radio Telescope revealed {sub-second pulsating emission features around 1\,GHz~\citep[][]{zhang23_ADleo_FAST,2025A&A...695A..95Z}}.
Figure.~\ref{fig:star_radio-mmQPP}a shows examples of radio bursts on AD\,Leo recorded by the upgraded Giant Metrewave Radio Telescope~\citep[uGMRT;][]{Gupta17_uGMRT} in the 500--850\,MHz range. The radio dynamic spectrum revealed strong quasi-periodic type III-like bursts followed by a type IV signature, making it the first report of solar-like burst types in a young M-dwarf. The band-averaged uGMRT light curve highlights the QPPs with multiple strong bursts marked from F1-4. 
Using modern sensitive interferometers like JVLA, LOFAR, ASKAP, etc., capable of wideband imaging observations, certain studies have reported radio burst fine structures, which could be classified as QPPs~\citep{Osten2006, Osten2008, villadsen19_Cohbursts_but_notypeII}. 

{The advancements in millimeter observations enabled by the Atacama Large Millimeter/submillimeter Array~\citep[ALMA;][]{Sven16_ALMA_science} and Northern Extended Millimetre Array~\citep[NOEMA;][]{Chenu16_NOEMA_techpaper} have also led to sub-second scale stellar flare studies in the 30--1000\,GHz band.} 
Figure~\ref{fig:star_radio-mmQPP}b shows a mm burst with second-scale QPPs and a characteristic positive spectral index around flare peaks, akin to some sub-THz solar flares~\citep[e.g.][]{silva1997,trottet08_2.2MeVline+200GHzsrc} and unlike other reported second-scale stellar bursts~\citep{MacGregor18_proxima_flares, Macgregor20_AUmicflare,Howard22_ProxCenFlare}. During the peak flare periods, significant variability in the linear polarization is also noted from the H and V polarization light curves.
In the sun, several mm burst QPPs have been observed with second-scale variability in flux, polarisation, and spectral index~\citep[e.g.,][]{kaufmann85_95GHzsolarflare, raulin03_fastmmQPPs_sun, Krucker13_mmSolarflareRev, Shen23_mmQPP_Xclassflare}. The frequency-rising bursts are often associated with particle acceleration in long flare loops~\citep[e.g.,][]{Fleishmann10_newMmSolFlareMech}, while QPP phenomena can result from various mechanisms discussed in Sec.~\ref{sec:mech}, depending on the event{\citep[e.g.,][]{Fleishmann10_newMmSolFlareMech,zaitsev14_plasmaemissMod_Mmpulses,2014ApJ...791...31K}.} Recently, \cite{Tandoi24_mmflarestars_SPT} released a catalog of stellar mm flares detected by the South Pole Telescope with minute-scale cadence, revealing QPPs in some events.


\section{QPP-oriented time-series analysis techniques}
\label{sec:techs}

Detection of QPPs in solar and stellar flares presents a major challenge for traditional time-series analysis because of their intrinsically non-stationary, short-lived, and often anharmonic nature. The conventional fast Fourier transform (FFT) technique assumes decomposition into stationary sinusoidal components and is consequently highly limited in application when the oscillation period and/or amplitude vary on timescales comparable to the flare duration. QPP signals typically last only a few oscillation cycles and undergo strong modulation driven by the evolving plasma parameters of the flaring region, rapid damping, or geometric changes in oscillating loops. As reviewed by \citet{2022SSRv..218....9A} {and \citet{2025NRvMP...5...21J}}, Fourier basis functions rarely match the true physical modes of the system, leading to power spreading across multiple harmonics and potentially spurious peaks in the spectrum. Noise, both white and coloured of physical or instrumental origin, together with the strong and often asymmetric flare trend, further distorts the spectral power distribution and complicates significance testing \citep[see e.g.][]{2017A&A...602A..47P, 2019PPCF...61a4024N}.

A partial improvement is obtained with the methods which provide oscillation power-frequency-time distributions and can therefore detect slowly evolving periodicities such as windowed Fourier transforms, wavelet transform and, in some cases, the Hilbert or Wigner--Ville (WV) transforms. These methods can capture QPP with weak non-stationarity, such as small frequency drifts. However, they remain fundamentally Fourier-based and therefore inherit the limitations of harmonic basis functions. They are especially sensitive to the broadband and intrinsically multi-scale nature of flare light curves. For example, the cross-term interference in the WV approach leads to the redistribution of power across frequency bands when multiple oscillatory components are present. Nevertheless, the method has been used successfully, for example, for the detection of QPP with modest frequency drift in microwave observations of a flare on AD~Leo \citep{2004AstL...30..319Z}.

To overcome the limitations inherent to Fourier-based approaches, empirical mode decomposition (EMD) has emerged as a powerful tool for QPP analysis \citep{2015A&A...574A..53K}. EMD imposes no assumption about basis functions: intrinsic mode functions (IMFs) are derived directly from the data, allowing extraction of locally defined, amplitude- and frequency-modulated oscillations. This adaptability makes EMD particularly well suited to QPP, which commonly display strong non-stationarity. It has been applied extensively to QPP (including those in non-thermal microwave, white light, and UV bands) in both solar and stellar flares \citep[e.g.][]{2018MNRAS.475.2842D, 2019MNRAS.482.5553J, 2020A&A...642A.195K}. Because noise can also generate spurious IMFs, EMD must be accompanied by rigorous statistical significance testing \citep{2016A&A...592A.153K}. Recent developments include the SunPy-affiliated SCOPE package\footnote{\url{https://github.com/Warwick-Solar/scope}}, designed to identify statistically significant oscillations extracted by EMD in the presence of coloured noise.

Another approach to QPP detection involves direct best-fitting of physically motivated models to the observed flare light curves with traditional least-square methods or a more advanced Bayesian Markov Chain Monte Carlo (MCMC) inference scheme. In particular, in the time domain, Bayesian MCMC enables fully probabilistic estimation of the characteristic flare and QPP timescales, including oscillation period/damping time variations \citep[e.g.][]{2020ApJ...905...70P, 2021ApJ...912...81K, 2025A&A...695L...4L}. It avoids assumptions inherent to spectral transforms but is necessarily model-dependent.
Another tool, AFINO\footnote{\url{https://aringlis.github.io/AFINO/}} \citep[Automated Flare Inference of Oscillations][]{2016ApJ...833..284I}, implements Bayesian MCMC best-fitting in the Fourier spectral domain, which compares competing power-spectral models to assess the presence of power enhancements caused by oscillatory components. Despite its systematic design, AFINO remains constrained by the fundamental assumptions and limitations of Fourier analysis and is therefore inherently conservative.

\begin{figure}[htb]
  \centering
  \includegraphics[width=\textwidth]{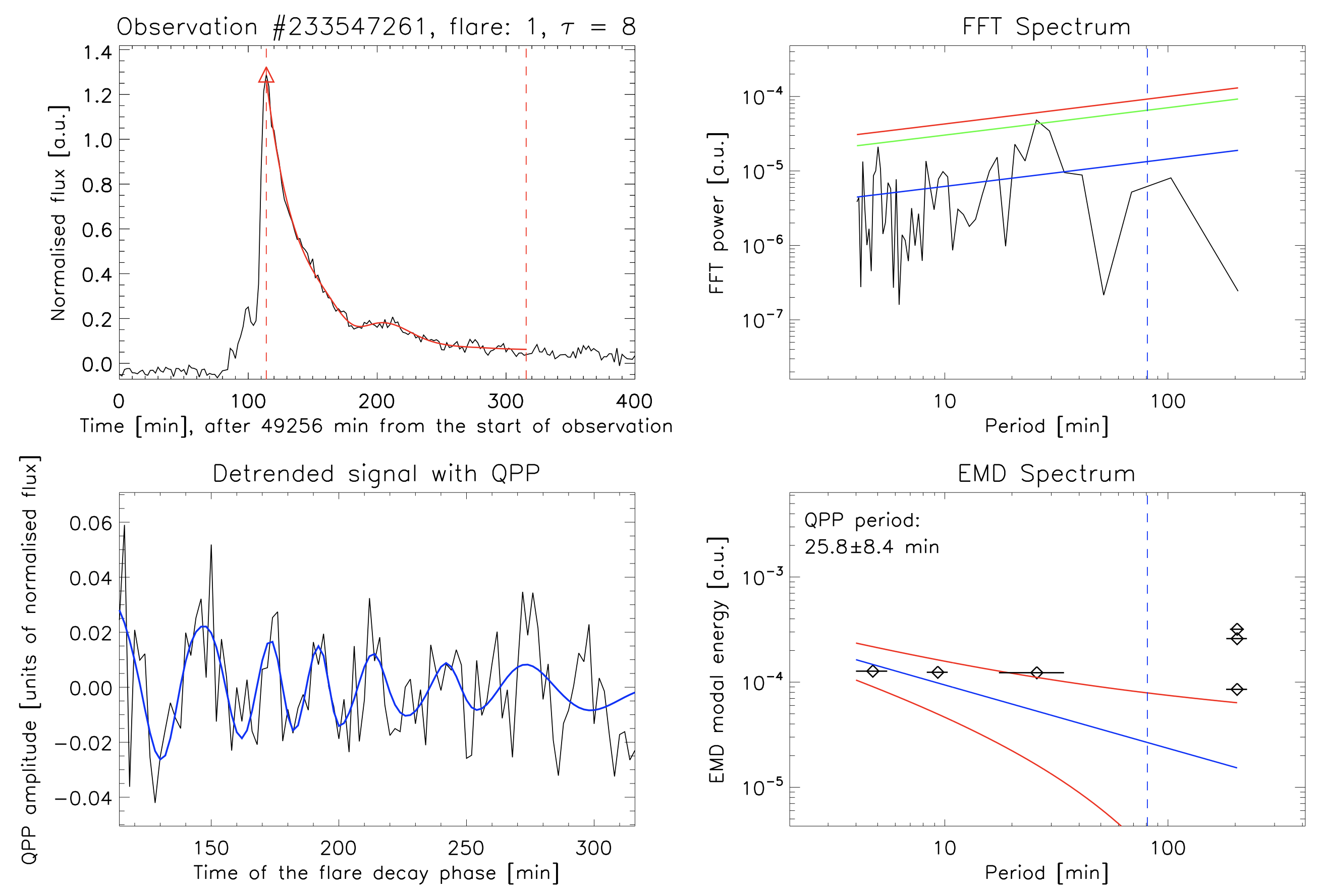}
   \caption{A QPP event with strong period drift in a white-light flare on TIC 233547261 star observed with TESS. Top left: flare light curve (black) with low-frequency trend (red); dashed lines mark the analysed interval. Bottom left: detrended signal (black) and the statistically significant EMD mode (blue). Top right: Fourier power spectrum of the detrended signal; the best-fitting noise model is shown in blue, with 1$\sigma$ and $2\sigma$ significance levels in green and red. The vertical dashed line marks the period equal to 0.4 of the analysed interval. Bottom right: EMD spectrum after trend removal; the expected power-law noise distribution (blue) and 95\% confidence levels (red) are shown. Error bars reflect the half-level widths of the global wavelet spectra for each mode. Figure credits: \cite{2021SoPh..296..162R}.}
   \label{fig-fft-emd}
\end{figure}

Because no single method is found to perform optimally for all QPP morphologies, the combined use of multiple, independent techniques seems to be the most robust practice. The large-scale comparative study of \citet{2019ApJS..244...44B} demonstrated that using two or more methods, typically one Fourier-based and one adaptive method such as EMD (see Fig.~\ref{fig-fft-emd}), significantly improves the detection reliability and minimises the probability of false positives across a wide range of QPP behaviours.

Advanced data analysis techniques are also required for the physical interpretation of QPP beyond their detection. In the radio band, the method of microwave imaging  spectroscopy \citep{2022Natur.606..674F} offers the capability of reconstructing time-dependent maps of plasma parameters in flaring active regions, linking observed QPP modulations directly to the underlying physical processes discussed in Sec.~\ref{sec:mech}. A complementary route involves studying empirically revealed scalings between QPP parameters, such as the dependence of damping time on oscillation period, which differ between proposed mechanisms. For example, resonantly-absorbing kink oscillations and standing slow waves damped by thermal conduction in coronal loops are expected to have linear and quadratic proportionality between damping time and period, respectively, which can be tested in QPP observations. Bayesian MCMC inference plays a central role in discriminating between competing models and comparing observations with multi-parametric theoretical predictions \citep[see e.g.][]{2023A&A...677A..23A}.

\section{Machine learning techniques in QPP study}
\label{sec:ml}

The explosive growth in observational facilities and the data provided by them demand the use of automated techniques to detect various phenomena, including QPPs. The standard transform-based data analysis techniques commonly used to detect QPPs in lightcurves \citep[see e.g.,][]{Broomhall2019} are limited due to non-stationary nature of QPPs and depend on the QPP type. In this regard, the machine learning (ML) techniques have the potential to supplement the traditional methods. In particular, ML open a prospect of using different data types for a detection/classification task (multimodal learning) extracting the useful information from as many data sources as possible, which is hard to achieve by using the traditional methods. Moreover, ML can be used to find dependencies in data, e.g., and to determine  data clusters. The latter has a great potential for classifying QPPs based on the data structure rather than the researcher's biases. Despite all the advantages of using ML in QPP studies, it is worth remembering that the majority of ML models, especially neural networks, are hardly explanable. This is why these models should be applied as preliminary data sifts to limit the amount  of data for traditional data analysis. 

\begin{figure}[htb]
  \centering
  \includegraphics[width=\textwidth]{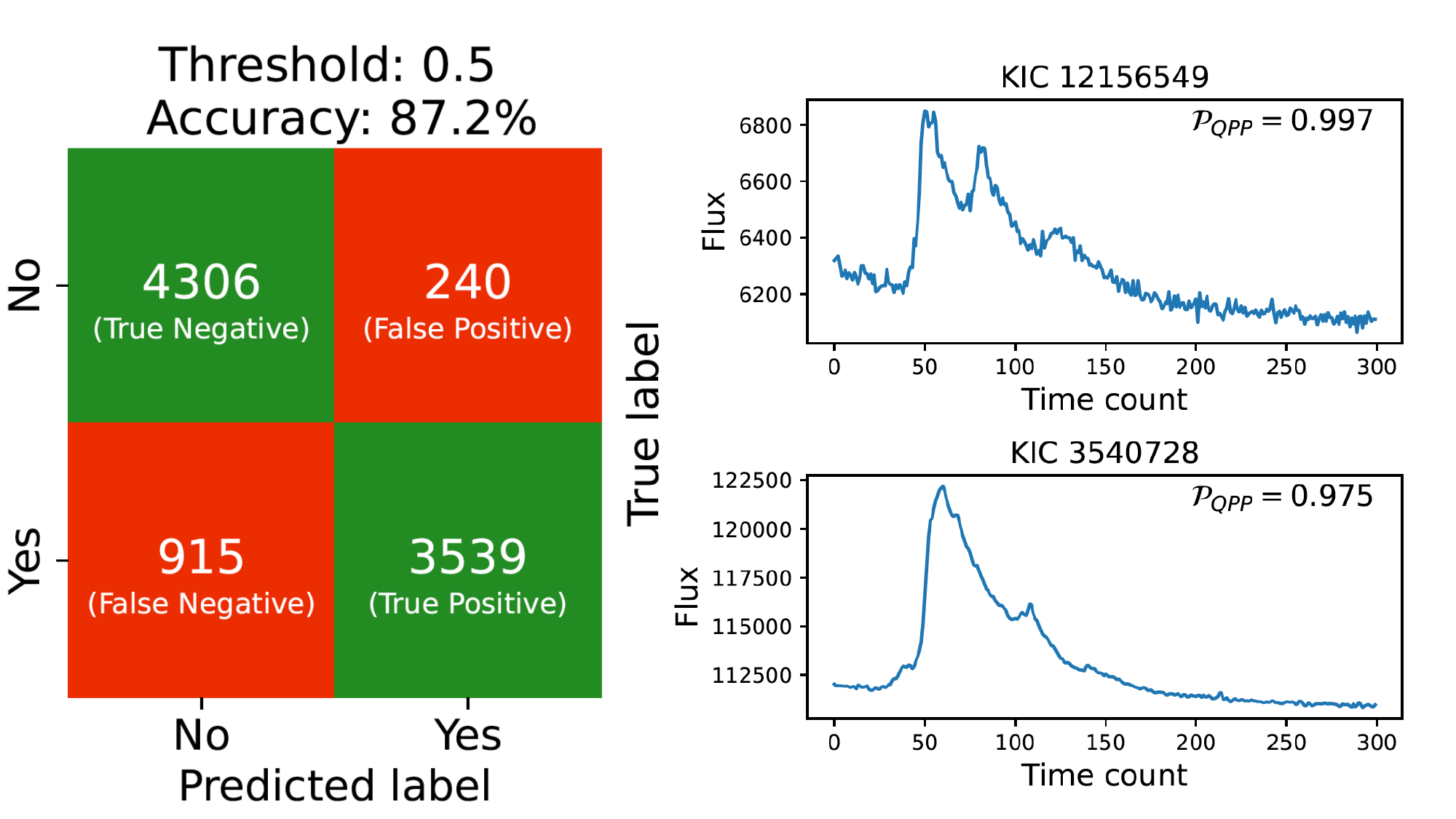}
   \caption{Left column: the confusssion matrix showing the FCN performance on the synthetic dataset. Right column: An example of the FCN result for two stellar flare lightcurves with QPP from \citep{2016MNRAS.459.3659P}. The numbers in the right upper corners show the FCN confidence in identifying the QPP. Figure credits: \cite{Belov2024}}
   \label{fig-ml}
\end{figure}

The first attempt to use ML in the QPP study was made by \citet{Belov2024}, where the Fully Convolution Network (FCN) architecture proposed by \citet{Wang2017}, was used to classify lightcurves, and determine whether they have exponentially decaying harmonic QPP or not. As this task requires a significant amount of data containing the exact type of QPP, the optimal approach is to generate the syntetic data to control the data quality. For this purpose, 90,000 synthetic flare lightcurves with and without QPP were generated. The network trained on this synthetic dataset demonstrated an accuracy
of 87.2\% (see the left column of Fig. \ref{fig-ml} for a confussion matrix). The right column of Fig. \ref{fig-ml} shows an example of the FCN responses for the lightcurves observed by Kepler where \citet{2016MNRAS.459.3659P} found decaying QPPs. In addition, the FCN was used to sieve the comprised of 2274 flare events observed by Kepler, resulting in a 7\% detection rate of QPPS of the assumed class, with a probability above 95\%. 

{Recently, \citet{2025ApJS..281...52W} applied the FCN developed by \citet{Belov2024} to identify a sample of 10,465 stellar flares detected with the Transiting Exoplanet Survey Satellite (TESS) exhibiting SUMER-type QPP phenomena. QPP are found to appear more frequently in M-type stars, in stars with lower effective temperatures, and in systems characterised by rapid rotation or short orbital periods, indicating clear correlations between the occurrence of QPPs and fundamental stellar parameters. No significant correlation is found between QPP periods and flare parameters such as energy, amplitude and duration. For the identification of flares, the authors used an ML search method based on a convolutional neural network developed by \citet{2020AJ....160..219F}. }

Future directions in QPP studies include the development of more sophisticated ML models, as well as the application of state-of-the-art models to detect, classify, and uncover relations in the QPP data. Another important avenue is the exploration of different data to enable multimodal QPP detection.This naturally expands the scope of data engineering: for example, additional diagnostic features such as the autocorrelation function may be incorporated, and raw lightcurves can be transformed into two-dimensional spectral representations for further analysis. Moreover, ML techniques applied to imaging data \citep{2025A&A...694A.237C,Belov2025} offer a route to link observed QPPs with specific wave modes, thus integrating temporal and spatial information in a unified framework. All of these steps will facilitate routine automated ML-powered QPP detection.

\section{Advances in QPP research with SKA precursors and pathfinders}
\label{sec:prec}

In this section, we present recent advances in QPP research using SKA precursors and pathfinders. 
The MWA facilitated high dynamic range snapshot spectroscopic imaging of solar active regions and quasi-steady emission regions across corona since the late 2010s. 
These studies required the development of novel imaging strategies and their implementation as automated pipelines, given the sheer volume of interferometric data ($\gtrsim$1\,TB/h) that MWA generates and the novel technical challenges involved in calibrating them \citep{Div17,Atul17,Mondal19_AIRCARS,Debo22_PAIRCARS_algo}.
Besides, automated analysis tools were required to identify, track, and model radio burst source morphologies across snapshot image spectral cubes generated by the imaging pipeline at sub-second and sub-MHz resolutions across 30\,MHz wide spectral bands \citep{Atul17,2023JApA...44...40O,Atul23_Rev}.
SPatially REsolved Dynamic Spectrum \citep[SPREDS;][]{Atul17,Atul23_Rev}, an extension of the traditional dynamic spectrum, is a powerful tool for exploring the evolution of specific regions of choice on the Sun. 
SPREDS allows performing source fitting in images made across every frequency-time bin and recording the fine spectro-temporal evolution of the fit parameters, to reveal the source morphological variability.
\begin{figure}[htb]
  \centering
  \includegraphics[width=0.9\textwidth,height=0.45\textheight]{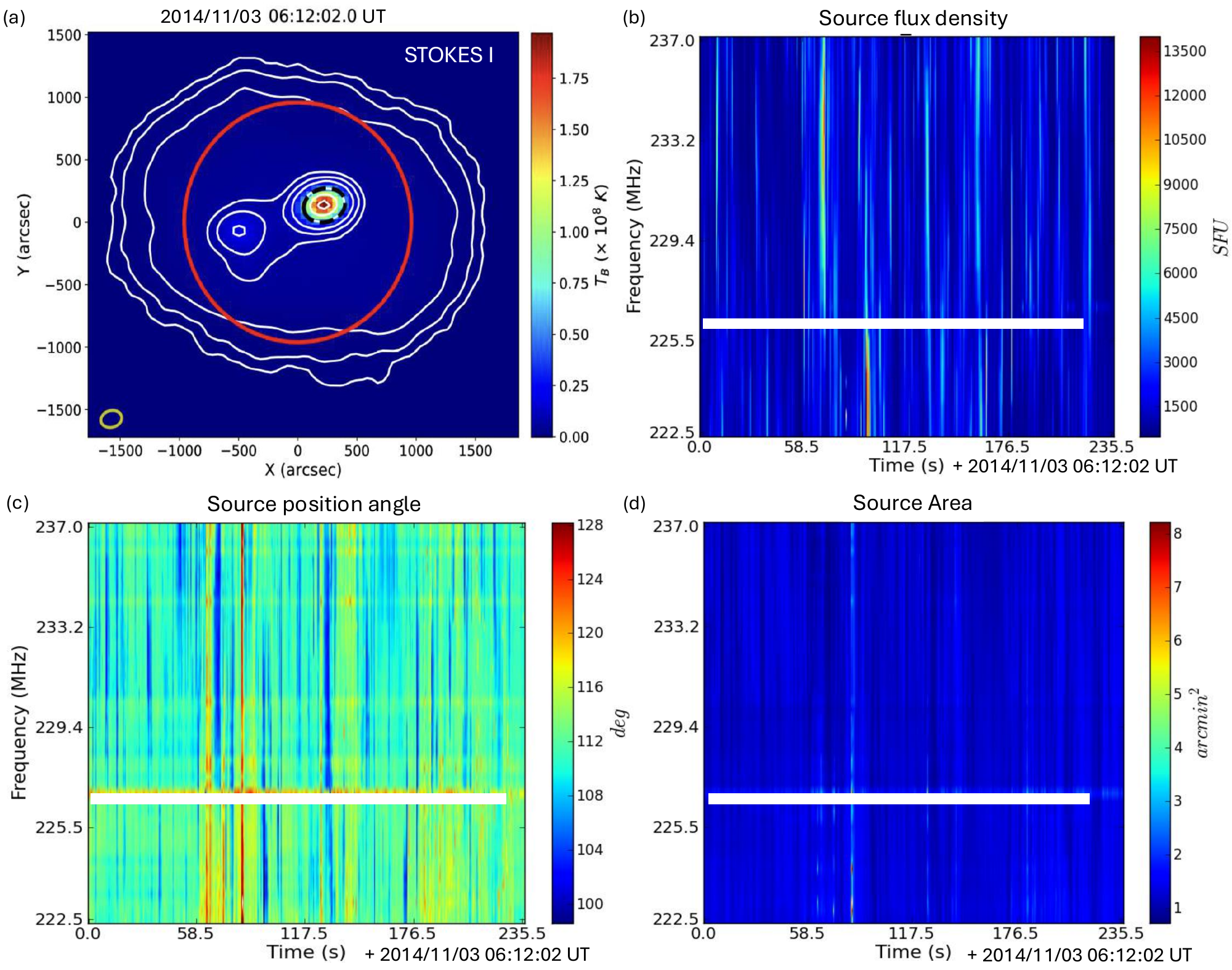}
   \caption{SPREDS example.(a) Snapshot spectroscopic image of the Sun at 229.36~MHz (0.5\,s, 160\,kHz averaging), with a bright type-I source highlighted by the bold black dotted contour. (b-d) SPREDS for the source flux density, area, and position angle derived by fitting the radio source with a 2D Gaussian function in every snapshot spectroscopic image made across 4 minutes and 15~MHz extent.}
   \label{fig0}
  \vspace{-0.5cm}
\end{figure}
Figure~\ref{fig0} shows an example of SPREDS for an active region, associated with a type-I noise storm. 
The radio source appeared to have a 2D Gaussian morphology (see Fig.~\ref{fig0}a). The data were fitted with a Gaussian source model and the integrated flux density, area, and position angle of the source were recoded for each time-frequency bin in which the data were imaged. The source area is estimated as the cross-sectional area of the ellipse that marks the full width at half maximum (FWHM) of the peak source flux density. In addition, the source centroids are also often explored to study the effects of radio-wave scattering and any inherent large scale dynamics \citep[e.g.,][]{Atul19_typeIIIQPP_turb}. 
Employing the novel imaging and analysis tools, namely SPREDS, solar active and quiescent emission regions were studied, exploring nanoflares to coronal mass ejections (CMEs) \citep[e.g.,][]{2017ApJ...851..151M,Rahman19_quietCorhole,Atul19_typeIIIQPP_turb,Atul19_microflareQPP,Mondal20_nanoflare,mondal20_CMEgyro,rohit20_QS,Atul21_dNN_vsht,2023ApJ...950..164K}.

Similar sub-second spectroscopic evolution of source sizes and flux density have also been performed with the SKA-Low pathfinder, LOw Frequency ARray~\citep[LOFAR;][]{vanHaarlem13_LOFAR} in the 30--80\,MHz band. These studies primarily explored the impact of coronal density fluctuations and other propagation effects on the observed radio source properties during weak to major solar flares and CMEs \citep[e.g.,][]{2017NatCo...8.1515K,Kontar19_scat, 2019ApJ...885..140Z, 2021A&A...645A..11M, 2025ApJ...978...73C, 2025A&A...700A.274K}.
Certain studies also explored the image plane sources of radio burst fine structures in the dynamic spectrum (e.g., type-III striae bursts), which are primarily caused by modulation of the radio emission by plasma density perturbations \citep[e.g.,][]{2016ApJ...826...78Y, sharykin2018_LOFAR_dnn_withtypIIIb, 2018ApJ...861...33K, 2021ApJ...917L..32C, 2025ApJ...978...73C}, for example, fast or slow magnetoacoustic waves \citep[e.g.,][]{2016ApJ...826...78Y}. These fine structures can be referred to as QPPs and may be linked with MHD wave processes in the emitting plasma, see \citep{2013ApJ...777..159Y, 2018ApJ...861...33K}. 


The high dynamic range MWA data and novel software tools facilitated prolonged exploration of active regions even during weak activity, irrespective of the occurrence of orders of magnitude stronger co-temporal radio bursts~\citep[see,][for an overview]{oberoi23_SKAInd}. This enabled the exploration of quasi-periodic variability in source morphology at sub-second scales from pre- to post flare phases during weak solar transients, which contribute largely to the coronal heating and particle acceleration than the major events~\citep{ash2012_flarestats}. 

\subsection{Radio `3D QPPs': insights and constraints to models}
\label{sec:3DQPP}

\begin{figure}[htb]
  \centering
  \includegraphics[width=0.96\textwidth,height=0.35\textheight]{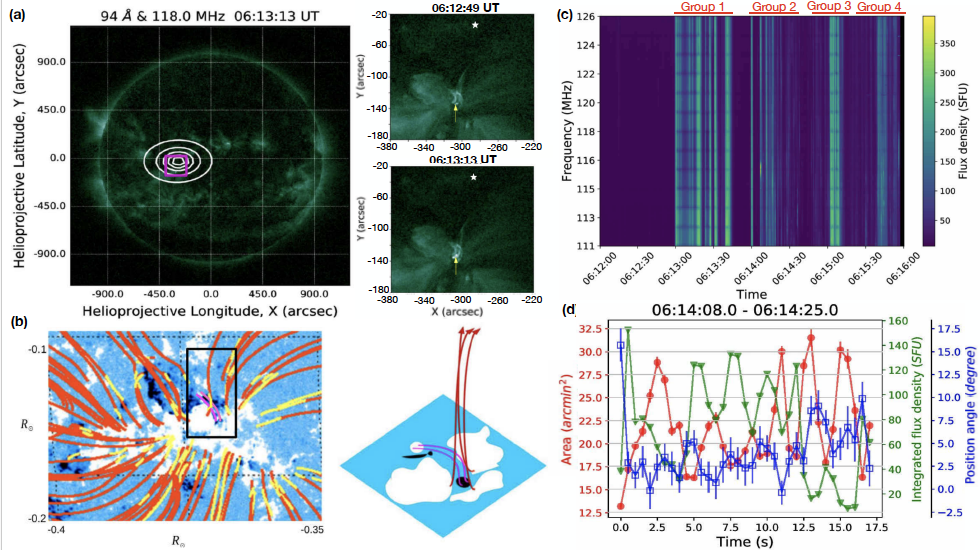}
   \caption{Weak jet event.(a) EUV image with radio source contours. Zoomed images of the active region (purple box in the full-disk map) before (06:12:49~UT) and after (06:13:13~UT) the start of the jet is shown beside, with the radio source centroid (star) and the reconnection site (arrow) marked. (b) NLFFF magnetic field model of the jet region (left) and a model-based schematic (right) of the open and closed field lines that reconnect to produce the jet. (c) Disk-integrated solar radio DS showing groups of broadband burst QPPs. (d) 3D QPPs: area, flux density, and position angle pulsations in the best-fit radio source function. Figure credits:\cite{Atul19_typeIIIQPP_turb,oberoi23_SKAInd}}
   \label{figWJ}
\end{figure}

\begin{figure}[htb]
  \centering
  \includegraphics[width=0.93\textwidth,height=0.35\textheight]{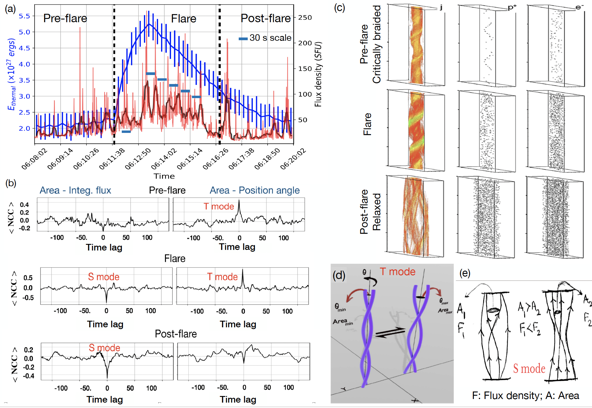}
   \caption{Microflare event. (a) Thermal energy evolution derived from EUV data analysis. Radio (200~MHz) source light curve at 0.5\,s resolution (red) and 10\,s running mean filtered light curve (black) showing the 30\,s QPPs. (b) Normalized cross-correlation (NCC) of the radio source position angle ($\theta$) and flux density (F) with area (A). 3D QPP mode evolves from T to S from pre-flare to the post-flare phase. (c) Simulation snapshots of magnetic field evolution of a critically braided coronal loop via microflare. The accelerated proton and electron density distribution along the loop imprints the field structure. (d-e) Schematic of origin of the `T' and `S' modes in 3D QPP sources envisaged with respect to the simulation model. Figure credits:\citep{Gordov12_MagRelax_loops,Atul19_microflareQPP,atul21_structQPPs}}
   \label{figMW}
   \vspace{-0.5cm}
\end{figure}

MWA imaging has revealed that QPPs previously identified only through flux modulation also show simultaneous changes in radio‐source morphology, introducing the concept of 3D QPPs \citep{Atul19_typeIIIQPP_turb,Atul19_microflareQPP,mondal21_longdurARQPPs}.
\citet{Atul19_typeIIIQPP_turb} found 2-s QPPs in the size and sky-plane orientation of a weak jet’s radio source, along with flux variability, giving the phenomenon a genuinely three-dimensional character. Figure~\ref{figWJ} shows the EUV jet, the associated radio-source contours, and the NLFFF reconstruction that indicates the interchange reconnection as the likely driver. The radio bursts originate from electron beams accelerated at the reconnection site and guided along open field lines; propagation and scattering effects explain small offsets between radio and EUV source locations. Type-III bursts appear in intermittent groups (Fig.~\ref{figWJ}c), and the correlated variability of area, orientation, and flux density (Fig.~\ref{figWJ}d) defines the 3D QPP behaviour. The observed 2-s morphological modulation is super-Alfvénic, implying that the reconnection process itself is periodically modulated by a local MHD mode.

Follow-up MWA studies found sub-minute 3D QPPs in both flaring and non-flaring intervals \citep{Atul19_microflareQPP,mondal21_longdurARQPPs}.
\citet{atul21_structQPPs} showed that these QPPs exhibit two characteristic modes: T (correlated size–orientation) and S (anti-correlated size–flux, sausage-like). Figure~\ref{figMW} highlights a microflare event where the radio source transitions from the T-mode in the pre-flare phase to the S-mode during and after the flare, reflecting a change in the dominant MHD mode as a critically braided magnetic field relaxes via bursty reconnection (Fig.~\ref{figMW}c–e). A 30-s modulation in the radio bursts agrees with the Alfvén speed inferred from the NLFFF geometry and DEM densities and matches the frequency extent of fine structures in the dynamic spectrum. This behaviour is consistent with a self-organised-criticality scenario in which small-scale instabilities trigger repeated reconnection episodes.

Only a few high-cadence MWA imaging–spectroscopy studies combine radio, EUV, X-ray data and magnetic modelling, but they demonstrate that 3D radio QPPs strongly constrain the underlying mechanisms and associated MHD modes. Further coordinated observations (MWA/LOFAR together with high-cadence EUV–X-ray spectroscopy) are needed to refine emission models.

Polarization calibration for large-N dipole arrays remains challenging, but new algorithms are improving this capability. Polarization measurements will help determine whether bursts are emitted at the fundamental or harmonic of the plasma frequency, improving local density and magnetic-field estimates, and improving the identification of QPP mechanisms.

\subsection{QPP detection in dynamic radio spectra with LOFAR}

\begin{figure}[htb]
  \centering
  \includegraphics[width=0.93\textwidth]{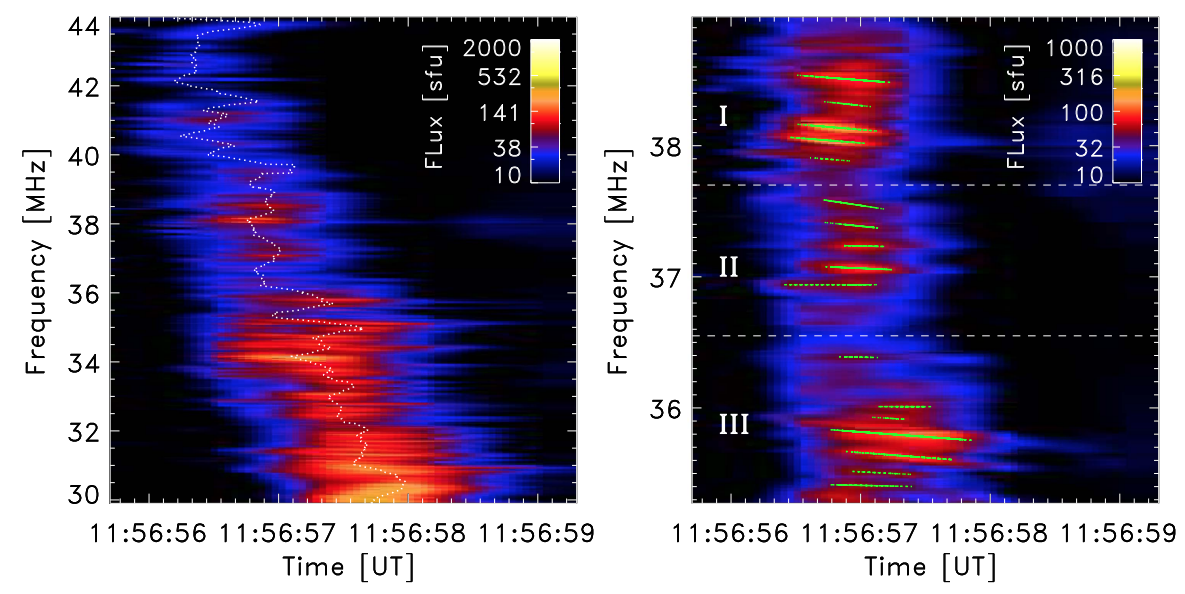}
   \caption{A 3-s interval of the dynamic spectrum, which is a Sun-integrated radio flux in the frequency–time plane, of a type III solar radio burst occurred on 2015 April 16 and observed by LOFAR. The left-hand and right-hand panels show the burst in the 30-–44~MHz and 35–-39~MHz frequency intervals, respectively. The white dotted line in the left-hand panel shows the \lq\lq spine\rq\rq\ of the burst, i.e., the instants of time of a maximum radio flux at each observational frequency. The straight green lines in the right-hand panel show fitting of the observed striae by a linear function. The regions of apparent clustering of the striae into three distinct groups are indicated as \lq\lq I\rq\rq, \lq\lq II\rq\rq, and \lq\lq III\rq\rq and separated by the horizontal dashed lines in the right-hand panel. Figure credits:\citep{2018ApJ...861...33K}.}
   \label{figLF}
   \vspace{-0.5cm}
\end{figure}

\citet{2017NatCo...8.1515K} presented LOFAR observations of a type~III solar radio burst from 2015 April 16 that reveal clear quasi-periodic striations in the dynamic spectrum at 35--39~MHz, see Figure~\ref{figLF}. These fine structures, detectable thanks to LOFAR's very high spectral resolution of 12~kHz, appear as narrow, drifting lanes of enhanced emission. By analysing the flux variation along the burst spine and converting the frequency to heliocentric height with the Newkirk density model, \citet{2018ApJ...861...33K} identified two distinct oscillatory components: a short-wavelength modulation of about 2~Mm and a longer-wavelength component near 12~Mm. The quasi-periodic behaviour is strongest between 1.63 and 1.69~$R_{\odot}$, above which the signal becomes dominated by noise or {turbulence \citep[e.g.][]{2018ApJ...856...73C}}.

\begin{figure}[htb]
  \centering
  \includegraphics[width=0.7\textwidth]{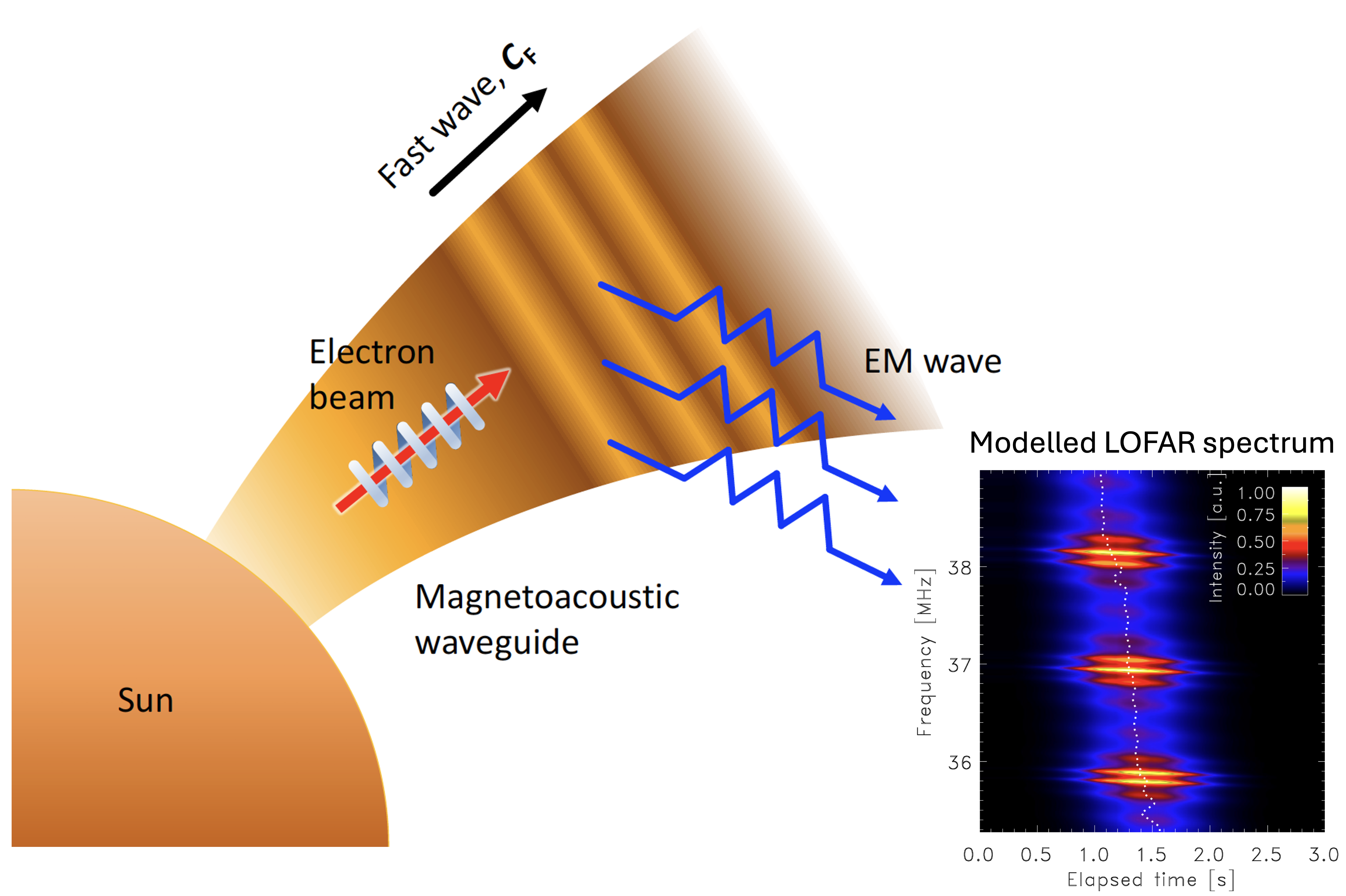}
   \caption{A schematic synopsis illustrating a qualitative scenario of the generation of quasi-periodic striation in the LOFAR dynamic spectrum of the type III burst by a propagating fast magnetoacoustic wave train.
   Figure credits: \citep{2018ApJ...861...33K}.}
   \label{fig-lofar-model}

\end{figure}

Wavelet analysis and measurements of the frequency drift \citep[see also][]{sharykin2018_LOFAR_dnn_withtypIIIb} showed that the short-wavelength striae propagate outward with a phase speed of approximately $657 \pm 114$~km~s$^{-1}$, corresponding to an oscillation period of about 3~s. These parameters strongly indicate the presence of a fast magnetoacoustic wave train guided by a coronal plasma non-uniformity aligned with open magnetic field lines, (see Fig.~\ref{fig-lofar-model}). Alfv\'en waves are excluded because their intrinsically non-collective behaviour and phase mixing would not produce coherent, spatially organised striations. The inferred periodicity and phase speed agree with previous detections of dispersive fast-mode wave trains in EUV and white-light observations, supporting the interpretation that a fast wave modulates the background plasma density.

The generation of the observed spectral fine structure was supported by a model in which the radio intensity is proportional to the amount of the emitting plasma with the electron plasma frequency, i.e., the density, corresponding to each narrow LOFAR frequency channel. It was found that a small-amplitude ($\sim 0.35$\%) density perturbation with a wavelength of roughly 2~Mm can reproduce the drifting, quasi-periodic striae. The observed grouping of striae into three broader clusters was reproduced by superimposing a second, longer-wavelength density modulation. The two-wave model successfully generated synthetic dynamic spectra closely resembling the observations.

The properties of the observed fast wave allow estimation of both the wave energy flux and the magnetic field strength in the emitting region. At a height of about 1.7~$R_{\odot}$, the wave energy flux is found to be modest, around 0.04~W~m$^{-2}$, which is at least an order of magnitude lower than the local radiative losses. Treating the measured phase speed as the fast-mode speed and combining it with density estimates derived from the plasma frequency yields an Alfvén speed of approximately 622~km~s$^{-1}$ and a corresponding magnetic field strength of about 1.1~G. This value is consistent with the  inverse-square decrease in the field strength with distance from the solar surface, assuming a magnetic field of about 3.1~G at the base of the corona.

Theoretical modelling of fast magnetoacoustic waves in plasma inhomogeneities with transversally non-uniform density profiles predicts characteristic time–frequency signatures, such as \lq\lq tadpole\rq\rq- or \lq\lq boomerang\rq\rq-like shapes in wavelet power spectra \citep{2021MNRAS.505.3505K}. Observational evidence for such fine spectral structures has been found, for example, in the decimetric radio burst at 973–-1025\,MHz analysed by \citet{2011SoPh..273..393M}. In that event, the wavelet spectrum of the radio emission intensity, modulated by a propagating fast wave train, evolves from a tadpole shape at higher frequencies (lower coronal heights) to a boomerang shape at lower frequencies (higher heights), with two well-pronounced arms. Quantitative analysis of such events can provide unique seismological diagnostics of the cross-field structuring of host active regions.

\subsection{State-of-the art in QPP research in the SKA-Mid band}\label{sec:QPP_GHz}

Unlike the SKA-Low band, the SKA-Mid band (0.35--15\,GHz) does not have high dynamic range large-N compact core imaging arrays like the MWA or LOFAR, essential for high dynamic range snapshot solar imaging. However, several studies have used state-of-the-art interferometers such as the Jansky Very Large Array (JVLA), the Extended Owens Valley Solar Array (EOVSA), the Siberian Radioheliograph \citep[SRH,][]{2017STP.....3a...3L}, and the long-standing Nan\c{c}ay Radio Heliograph (NRH) for QPP studies in the SKA-Mid band. In addition, there have also been imaging studies using the Nobeyama Radioheliograph (NoRH) at 17\,GHz, close to the SKA-Mid band. The majority of these imaging studies of QPPs has been focused on strong flares close to the flare peak and decay phases~\citep[see,][for an overview]{2009SSRv..149..119N}.

The brightness temperatures of the two sources, both at 3.4~GHz and 8.4~GHz, associated, respectively, with optically-thick and optically-thin emissions, increase and decrease repeatedly and in phase.
It was suggested that the detected QPP were connected with repeatitive magnetic reconnection at the flare current sheet.

\citet{2024A&A...684A.215C} utilised high-resolution microwave imaging with EOVSA together with hard X-ray observations to pinpoint the spatial and temporal characteristics of 7--35~s QPP during an X1.3 solar flare. The authors identified distinct microwave sources associated with accelerated electrons in the corona and showed that the pulsations were synchronised across hard X-rays and microwaves, indicating a common origin. By examining source morphology, spectral behaviour and timing, it was concluded that the QPP are best explained by periodic modulation of the electron acceleration and/or trapping region, and proposed that repeatative reconnection or oscillatory magnetic topology were likely responsible.

QPP have also been detected with SRH in the 4--8~GHz band. The microwave emission displayed a broadband gyrosynchrotron spectrum modulated with a periodicity of approximately 30~s and originating from a compact, loop-like source at a height of approximately 31~Mm \citep{2021SoPh..296..185K}. The QPP was associated with a standing sausage oscillation in the flaring loop.  

\section{Expected advances with SKA-Low and SKA-Mid}
\label{sec:exp}

The SKA Low and Mid arrays will observe in the 0.05--15\,GHz band that probes plasma and magnetic field variability across a wide range of coronal heights, from the low corona to a height of about 2--3\Rsun, depending on the emission mechanism \citep{Zucca01.2026.SKA, Kontar01.2026.SKA, Kumari01.2026.SKA, Morosan01.2026.SKA}. 
When coherent plasma emission generally dominates during magnetic activity in the SKA-low band, SKA-mid will detect emission from both coherent and incoherent processes (see Sec.~\ref{sec:QPP_GHz}), providing diagnostics of non-thermal electron population and magnetic field strengths across low-coronal heights.
The arrays will enable sensitive sub-second sub-MHz snapshot spectro-polarimetric solar and stellar imaging observations across the wide range of coronal heights, essential for QPP research. 
Solar and stellar studies, primarily based on high-resolution disk-integrated dynamic spectra, have shown that emission variability at such scales exists that informs on MHD scales at the active region.  
The large-N architecture and enhanced sensitivity of the SKA-Mid and Low arrays will enable high-fidelity solar imaging required for QPP studies. The arrays will offer significantly higher sensitivities in the respective spectral bands of operation compared to current state-of-the-art facilities, due to their relatively much higher collecting area (A$_e$) and lower system temperature (T$_{sys}$). In addition, the much longer baselines will allow for an order of magnitude better angular resolution relative to the existing arrays operating in the same spectral bands.

{\bf SKA-Mid} will have more than 133 antennas enabling high dynamic range imaging at about 0.14\,s and 13\,kHz resolution.
For comparison, at 1.4\,GHz, SKA-Mid will offer $\sim$ 5 times better sensitivity in the AA* (AA4) configuration than the current state-of-the-art instrument, MeerKAT. 
Besides, SKA-Mid's angular resolution of about 0.05$^{\prime\prime}$ $\times$ 0.04$^{\prime\prime}$ in the image plane, which is a significant improvement compared to the few arcsec resolution offered by the solar imaging mode of the JVLA\footnote{\href{https://science.nrao.edu/facilities/vla/docs/manuals/obsguide/modes/solar}{https://science.nrao.edu/facilities/vla/docs/manuals/obsguide/modes/solar}}.

{\bf SKA-Low} will also significantly enhance the imaging dynamic range owing to the planned 512 stations operating as phased-up antenna elements at similar time-frequency resolution as the SKA-Mid. This is a significant improvement compared to the current imaging capability of MWA with 256 antenna elements \citep{Wayth18_MWA_phaseII,Morrison23_MWAXcorr}.
The SKA-Mid and SKA-Low in the upcoming AA* configuration and the later AA4 configurations can hence let us explore sub-second-scale variability in the QPP sources across the 0.35--15\,GHz band in a much better way with about an order of magnitude better spatial resolution for events ranging from nanoflares to major flares and CMEs.

\subsection{Prospects for high cadence multi-waveband imaging exploration of QPPs}
Recent developments in EUV and soft X-ray (SXR) observations have enabled regular high cadence observations of the corona.
The Extreme Ultraviolet Imager~\citep[EUI;][]{Rochus_SolO_EUI} onboard
the Solar Orbiter~\citep[SolO;][]{SolO_2020} is able to provide high cadence ($\sim$s) and high resolution images of the solar atmosphere in  the 174~{\AA} passband. Interestingly, recent high-cadence EUV observations by \cite{Petrova23_MHDwavesin30sQPPs} using EUI onboard SolO revealed the presence of inhomogeneities (which they interpret as kinks) at a scale of $\sim$11\, Mm along a long coronal strand that showed 30\,s QPPs in EUV emissivity.
A recent survey of EUV brightenings carried out by EUI have already found statistical samples of QPPs in flares, active regions, and in quiet sun ranging from 5 -500\,s timescales~\citep{Lim_QPP_2025}.

Similarly, Interface Region Imaging Spectrograph~\citep[IRIS;][]{Depontieu14_IRIS} observations of active regions have also revealed second-scale variability at the footpoints of active regions in the transition region and chromospheric heights~\citep[e.g.,][]{Testa14_ARmossheating,Testa20_mossvariablityIRIS}. 
A recent study combining high spectral and spatial resolution IRIS data with simultaneous observations of the $\approx$1~MK corona of an active region with the High-resolution Coronal Imager \citep[HiC2.1;][]{2019SoPh..294..174R} sounding rocket, revealed clear second-scale spatio-temporal cross-correlations between heating in the chromosphere and the corona~\citep{bose24_ARplageheating}. 
This study highlighted a tight dynamical link and coupling between the heating across atmospheric layers, and the potential role of QPPs in the underlying processes. 

EUV observations have demonstrated a possible link of chromospheric oscillations with oscillatory wave processes in coronal active regions. In particular, 3-min oscillations in sunspot umbrae appear to be drivers of slow magnetoacooustic waves which are detected as almost monochromatic perturbations of EUV intensity propagating upward along coronal plasma feathers \cite[see, e.g.,][]{2009SSRv..149...65D, 2021SSRv..217...76B}. These waves indicate the inter-connectivity of various layers of the solar atmosphere. Furthermore, the leakage of chromospheric 3-min oscillations into the corona has been found to produce 3-min QPP of microwave emission from flaring sites \citep{2009A&A...505..791S}. It has recently been established that coronal 3-min waves propagating along different magnetic flux tubes anchored at different locations in the umbra have statistically significant oscillation periods \citep{2025MNRAS.536.3192M}. This finding allows for a unique identification of individial waveguiding flux tubes in multi-instrumental quasi-stereoscopic data, creating a ground for the seismological determination of the local magnetic field direction. This information can be used to validate and constrain NLFFF codes. The high-cadence observational capabilities of the SKA-Mid and EUV instruments will allow resolving fine spectral features in coronal slow waves, which is crucial for the seismological diagnostics.  

\begin{figure}[htb]
  \centering
  \includegraphics[width=0.70\textwidth]{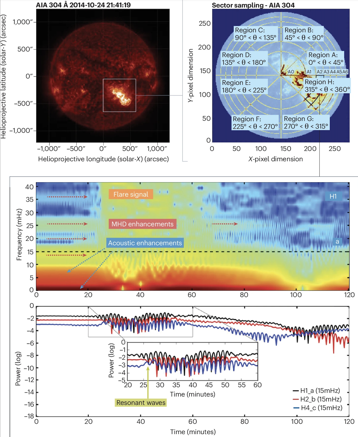}
   \caption{Top left: SDO AIA 304{\AA} image of the X3.1 flare occurring 24 October 2014, around active region AR 12192. Top right: to facilitate the detection of wave behaviour, the flaring region is uniformly partitioned into 45 degree angular sectors. Each sector is further subdivided into seven discrete regions, resulting in a total of 56 regions, to enable precise sampling and analysis. Middle: a sample spectrogram of the sectored region H1, encompassing the main body of the flare, reveals both MHD and acoustic enhancements occurring before and after flare ignition, approximately 25 minutes in. These enhancements are attributed to Alfv\'en and MHD waves, as well as acoustic waves propagating through the chromospheric region, which subsequently transform into magnetoacoustic waves travelling in the corona near an active region associated with a solar flare. Bottom: we also observe an increase in the amplitude of the waves, indicating `resonance' behaviour within the flare, which may result from an unknown process, and could be indicative of oscillatory reconnection. Figure credits: \citet{2025NatAs...9..760C}} 
   \label{fig-pmm}
\end{figure}

The application of the 3D \emph{Frequency-filtered Amplitude Movies} \citep[FFAMS,][]{2013ApJ...765...15J} method, also known as Power Map Movies (PMMs), to the H$\alpha$ data, observed with the United States Air Force/National Solar Observatory Improved Solar Observing Optical Network (ISOON) telescope \citep{2013ApJ...765...15J} revealed the variation of the chromospheric acoustic power around solar flare sites. This technique was later incorporated by \citet{2016AJ....152...81M} to further analyze the wave excitation around solar flares. This analysis was expanded by extracting information from the PMMs in the form of periodograms, using GONG H$\alpha$ detectors. The further evolution of the PMM technique is proposed by \citet{2025NatAs...9..760C}, in which spectrograms extracted from sectored regions detect wave power enhancements before the flare onset and could be indicators of oscillatory reconnection, see Figure~\ref{fig-pmm}. Furthermore, the use of more sophisticated techniques for the analysis of 3D data cubes, such as the Pixelised Wavelet Filtering \cite{2008SoPh..248..395S} may further strengthen those findings.  

\subsection{Coordinated observation possibilities involving SKA}
This section discusses some future EUV missions which will revolutionise the field of high cadence imaging spectroscopy sensitive to heights within upper chromosphere and corona. These instruments in combination with non-thermal diagnostics from SKA are well poised to make major leaps in the understanding of QPP physics. 

{\bf The Multi-slit Solar Explorer} \citep[MUSE;][]{DePontieu_2020_MUSE} is a multi-slit spectrograph designed to study the solar atmosphere through spectroscopy in three bands centered around Fe IX 171~{\AA}, Fe XV 284~{\AA}, Fe XIX 108~{\AA} and Fe XXI 108~{\AA}. These lines are sensitive to plasma at temperatures of $\approx$ 0.7, 2, 10, and 12~MK respectively. The instrument performs spectroscopy across 35 slits spanning  $\sim150''\times170''$, with a spatial resolution ($0.167''$ along the slits and $0.4''$ across the slits), with cadences of 12 seconds. MUSE would resolve the velocities in the Fe~IX and Fe~XV lines with a resolution of $\sim5$ km/s, and $\sim30$ km/s in Fe~XIX and Fe~XXI, with a line width resolution of $\sim10$ km/s ($\sim30$ km/s) in Fe~IX and Fe~XV (Fe~XIX and Fe~XXI). These capabilities allow MUSE to effectively ``freeze'' the evolution of dynamic coronal plasma, capturing temporal variations that are the hallmark of QPPs \citep{Cheung_2022_MUSE2Flares}. The MUSE observations allow for diagnostics in a wide range of coronal temperatures, providing intensities, velocities and non-thermal line widths. The multi-slit approach enables simultaneous observations covering the flare loop tops thereby capturing both global and local dynamics of QPPs. Furthermore, MUSE also carries two context imagers onboard, with passbands centered around 195~{\AA} and 304~{\AA} at a spatial resolution of $\sim0.33''$ and going upto a time cadence of $\sim4$s. This provides supporting data on the global context of the flare ribbons and loop dynamics, enhancing the identification and analysis of QPP signatures in the solar atmosphere. 

{\bf The Extreme Ultraviolet High-Throughput Spectroscopic Telescope}~\citep[EUVST;][]{EUVST} is a solar telescope designed to observe the Sun's atmosphere with very high spatial (0.4 arcsec) and temporal (fastest will be 0.5 s) resolution across a field of view of $\sim100''\times100''$. It will observe a wide range of temperatures simultaneously, from the chromosphere at 0.02 MK, up to heated solar flare plasmas at 15 MK. The wavelength ranges of EUVST is split into the short wavelength at 170--210~\AA{} and the long wavelength at 690--1275~\AA{}.  The science goals of EUVST that target QPPs is to understand how the solar atmosphere becomes unstable, releasing the energy that drives solar flares and eruptions. The capability of EUVST will specialise in observing the atmospheric response (density, temperature, Doppler flows, non-thermal velocities) to energetic particles that drive QPPs from a huge range of altitudes that correspond to the wide temperature coverage.  There is scope for synergistic observations with the SKA, particularly during flares, because of the large frequency coverage of the SKA capturing radio emission from the base of the corona out to at least one solar radii in altitude.

{\bf The Solar Ultraviolet Imaging Telescope}~\citep[SUIT;][]{SUIT_Aditya, suit_calib} onboard Aditya-L1~\citep{AdityaL1} provides full-disc imaging observations in 11 passbands, spanning the photosphere and chromosphere. The instrument is set to provide observations at a pixel size of 0.7'', at a nominal time cadence of 20~s. However, SUIT can also reach a 4~s cadence in specific regions. Observations of flares with SUIT show propagation of multi-thermal blobs, which show association with pulsations in hard X-ray observations~\citep{suit_flare}. 

The QPP study will also benefit from observations from the Sun–Earth L4 and L5 Lagrange points, which could provide a crucial off-axis perspective that greatly reduces the projection effects inherent in Earth-based viewing. From these locations, the 3D geometry of flaring loop systems—loop lengths, inclinations, apex heights, and current-sheet structures, can be determined with far greater fidelity, enabling more accurate diagnostics of QPP properties. The L4/L5 vantage also allows spatial disentanglement of overlapping post-flare arcades and plasmoid-rich current sheets, thereby facilitating precise localization of the regions where QPP originate. In addition, side-on viewing supports reliable tracking of propagating QPP wave trains and periodic signatures in reconnection outflows and CME-driven shocks. By providing continuous projection-free coronal imaging during flares, \textbf{L4} \citep{2023JKAS...56..263C} and \textbf{L5} \citep{2022cosp...44.3544P} missions will form an essential complement to the high-sensitivity radio measurements from SKA, jointly enabling more definitive discrimination among the physical mechanisms that produce QPP in solar flares.

\textbf{The Solar Particle Acceleration, Radiation and Kinetics (SPARK)} mission concept is currently under consideration in the ESA M8 call.  SPARK is planned to contain EUV imagers that will tackle the lack of spatially resolved high-cadence observations that has prevented a definitive  explanation of QPPs. SPARK will, for the first time, provide the high-resolution diagnostics required to determine the origin of QPPs. The SPARK EUV imager (HiFI) will capture sub-second EUV evolution from coronal plasma in the flaring region.  The EUV imaging spectrometer (SISA) will provide plasma Doppler from 1--15 MK plasma in the 18--26 nm wavelengths.  The X-ray/gamma-ray spectrometer (LISSAN) will identify modulated emission from accelerated electrons through sub-second X-ray spectra around 10's keV, and potentially long-period (> 1 minute) QPPs present in the gamma-ray signature of accelerated ions. These data will distinguish between wave-driven modulation and episodic reconnection, revealing whether QPPs are a fundamental signature of the flare energy release process.  If selected, SPARK will launch in 2041 and complement the SKA diagnostics of QPPs by combining  EUV and X-ray/gamma QPP signatures with radio emission from accelerated particles in solar flares.

\section{Conclusions}
\label{sec:conc}

Quasi-periodic pulsations (QPPs) in solar and stellar flares are attracting growing attention as a missing element of the standard flare model, a powerful diagnostic of energy release and particle acceleration, and a universal process that enables exploitation of the solar–stellar analogy. From radio and EUV to X-rays, gamma rays, and white light, QPPs reveal the highly structured, multi-scale nature of flare dynamics, reflecting a combination of oscillatory processes, episodic acts of spontaneous and driven magnetic reconnection, and various plasma instabilities. However, despite significant observational progress, detections remain limited by instrumental sensitivity, cadence, and incomplete spatial coverage—particularly for weak flares and for stellar events where spatially resolved information is unavailable.

The results summarised in this chapter highlight how modern high-cadence broadband radio imaging spectroscopy with the SKA will significantly advance QPP studies. The synergy between high-cadence radio imaging, EUV and X-ray diagnostics, DEM-based plasma characterisation, NLFFF magnetic reconstruction and state-of-the-art theoretical modelling allows QPP research to place quantitative constraints on local plasma parameters and characteristic spatial and temporal scales. Such constrained modelling is essential for distinguishing between competing physical interpretations—for example, oscillatory reconnection, slow or fast magnetoacoustic modes, or self-organised-criticality–driven energy release. Of particular importance is the unprecedented sensitivity of SKA, which opens a unique opportunity to study microflares and nanoflares, including their QPP signatures. This research direction may fundamentally transform our understanding of the long-standing problem of solar and stellar coronal heating.

At the same time, the rapid growth in data volumes and morphological diversity necessitates robust automated detection and classification frameworks. This chapter summarises the strengths and limitations of Fourier, empirical mode decomposition, and Bayesian MCMC methods, and highlights the increasing role of machine-learning approaches for identifying QPP signatures across large datasets. Machine learning—applied to both light curves and imaging data—is emerging as a powerful tool for linking observed QPP behaviour with theoretical models. The design and application of advanced data-analysis techniques will be essential for exploiting the full potential of SKA in QPP research.

Overall, the phenomenon of QPP offers a unique, and still underexplored, source of information about the physical conditions and dynamic processes in solar and stellar atmospheres. Through the transfer of MHD coronal seismology from solar to stellar physics, and through effective use of the solar–stellar analogy, QPP studies hold the promise of delivering transformative insights across the broader field of stellar activity.

\bibliographystyle{abbrvnat-maxbibnames4}
\bibliography{qpp_refs} 

\end{document}